\documentclass[runningheads]{llncs}
\usepackage{graphicx}
\usepackage{hyperref}

\makeatletter
\renewcommand\subsubsection{\@startsection{subsubsection}{3}{\z@}%
                       {-18\p@ \@plus -4\p@ \@minus -4\p@}%
                       {0.5em \@plus 0.22em \@minus 0.1em}%
                       {\normalfont\normalsize\bfseries\boldmath}}
\makeatother
\usepackage{booktabs}
\usepackage{multirow} 
\usepackage{makecell} 
\usepackage{textcomp}
\usepackage{gensymb} 
\usepackage{stmaryrd} 

\usepackage{cite}
\usepackage{sidecap} 
\usepackage{caption}
\usepackage{subcaption}
\usepackage{ulem} 
\usepackage{pdfpages}
\usepackage{subcaption} 

\usepackage{setspace}
\usepackage{amssymb}
\usepackage{amsmath}

\begin{document}

\title{Multi-dimensional topological loss for cortical plate segmentation in fetal brain MRI}
\titlerunning{Multi-dimensional topological loss for CP segmentation in fetal MRI}

\author{Priscille de Dumast\thanks{Corresponding author, priscille.guerrierdedumast@unil.ch} \inst{1,2} \and
Hamza Kebiri\inst{1,2} \and
Vincent Dunet\inst{1} \and \\
Mériam Koob\inst{1} \and
Meritxell Bach Cuadra\inst{2,1}}
\authorrunning{P. de Dumast et al.}

\institute{Department of Diagnostic and Interventional Radiology,\\ Lausanne University Hospital and University of Lausanne, Lausanne, Switzerland \and
CIBM Centre d'Imagerie BioMédicale, Lausanne, Switzerland}

\maketitle

\begin{abstract}
The fetal cortical plate (CP) undergoes drastic morphological changes during the \textit{in utero} development. Therefore, CP growth and folding patterns are key indicator in the assessment of the brain development and maturation. Magnetic resonance imaging (MRI) offers specific insights for the analysis of quantitative imaging biomarkers. Nonetheless, accurate and, more importantly, topologically correct MR image segmentation remains the key baseline to such analysis. 
In this study, we propose a deep learning segmentation framework for automatic and morphologically consistent segmentation of the CP in fetal brain MRI. 
Our contribution is two fold. First, we generalized a multi-dimensional topological loss function in order to enhance the topological accuracy.
Second, we introduced hole ratio, a new topology-based validation measure that quantifies the size of the topological defects taking into account the size of the structure of interest. Using two publicly available datasets, we quantitatively evaluated our proposed method based on three complementary metrics which are overlap-, distance- and topology-based on 27 fetal brains. Our results evidence that our topology-integrative framework outperforms state-of-the-art training loss functions on super-resolution reconstructed clinical MRI, not only in shape correctness but also in the classical evaluation metrics (mean$\pm$std: Dice similarity coefficient of 0.85$\pm$0.01, average symmetric surface distance of 0.19$\pm$0.01 mm and hole ratio of 0.06$\pm$0.03). Furthermore, results on additional 31 out-of-domain SR reconstructions from clinical acquisitions were qualitatively assessed by three experts. The experts' consensus ranked our \textit{TopoCP} method as the best segmentation in 100\% of the cases with a high inter-expert agreement. Overall, both quantitative and qualitative results, on a wide range of gestational ages and number of cases, support the generalizability and added value of our topology-guided framework for fetal CP segmentation.

\end{abstract}

\newpage

\section{Introduction}
\label{sec:introduction}

\subsection{Clinical context}
\label{sec:relatedWork}

During \textit{in utero} development, the human fetal cortical plate (CP) that is the future cortex undergoes drastic changes \cite{tierney_brain_2009}.
Indeed, the brain moves from a smooth surface at 10 weeks of gestation to a convoluted one at 35 weeks thanks to the appearance of the cortical gyrifications (see Figure~\ref{fig:Fig_morpho_f_GA}). Jointly, the surface area and the volume of the future cortex are respectively increased 50 and 40 times during the 2nd and 3rd trimesters of pregnancy~\cite{vasung_quantitative_2016,vasung_spatiotemporal_2020}.

Cortical folding consists in the chronologic appearance of primary, secondary and lastly tertiary sulci~\cite{garcia_mechanics_2018}. Nearly all gyri are in place at birth, even though the gyrification process carries on afterward \cite{lenroot_brain_2006}.
Cortical gyrification is considered to be a relevant marker of fetal brain maturation, as the chronological sequence of appearance of sulci is well known during the fetal period~\cite{garel_fetal_2003}.
Conversely, abnormal fetal sulcation and gyration indicate disruption of one of the three main fetal stages of normal cortical development (i.e. cell proliferation, migration and cortical organization)~\cite{barkovich_developmental_2005,leibovitz_fetal_2022}.

Altered cerebral cortex formation, induced either by genetic mutations, vascular injuries, metabolic diseases, fetal infection or teratogenic causes, may lead to malformations of cortical development. An updated classification of this group of heterogenic disorders has been recently published in a consensus statement~\cite{severino_definitions_2020}. Those rare disorders usually manifest with developmental delay, seizures, and motor and sensory deficits~\cite{leventer_malformations_2008}. 
Given the consequences of abnormal brain gyrification, early diagnosis is crucial, for which the analysis of cortical maturation is an asset. 

Advanced neuroimaging techniques can greatly benefit the characterization of neurotypical and pathological fetal brain development. 
Ultrasound (US) is the first prenatal imaging modality for fetal screening. Unfortunately, its sensitivity to the surrounding maternal tissues easily alters the image quality. Conversely, magnetic resonance imaging (MRI) does not share these limitations and hence comes as a key complementary imaging tool to look for additional information in equivocal situation or to confirm or rule out eventual US pathological findings \cite{salomon_third-trimester_2006,griffiths_prospective_2010,prayer_isuog_2017,griffiths_mri_2019}.
MRI is a non-invasive and reliable imaging method for the monitoring and follow-up of fetal brain development that relies on the tissue properties to generate image contrasts~\cite{gholipour_fetal_2014}. MRI is a valuable technique highly suitable for evaluating fetal brain morphometry and connectivity in neurotypical and pathological cases~\cite{garel_mri_2004, prayer_isuog_2017}. In fetal brain MRI, T2-weighted (T2w) sequences evidence soft-tissue contrast. In this respect, they are essentially used to assess morphological and structural development \cite{gholipour_fetal_2014}.

In clinical practice, fast MR sequences are run in order to “freeze” the unpredictable fetal motion in the plane of acquisition \cite{prayer_isuog_2017}. 
While the series presents an excellent in-plane resolution (1 to 2 $mm^2$), slice thickness has a sub-optimal resolution (3 to 5 mm) \cite{gholipour_fetal_2014}. In a clinical examination, several low-resolution (LR) series are acquired in three orthogonal planes to offer complementary structural information from the three dimensions. However, the strong anisotropy of the volumes and the remaining inter-slice motion hampers a correct realignment of the images in the anatomical planes and corrupts any three-dimensional based measurements. 
Over the last decades, novel advanced image processing algorithms, based on super-resolution (SR) methods, were developed for the reconstruction of 3D volumes \cite{rousseau_registration-based_2006, gholipour_robust_2010, kuklisova-murgasova_reconstruction_2012, tourbier_efficient_2015, ebner_automated_2020}. The underlying reconstruction principle is to combine a set of clinical LR acquisitions into a single high-resolution isotropic motion-free volume~\cite{uus_retrospective_2022}. In addition to improved visualization, the availability of SR reconstructed volumes opens up to more accurate 3D-based quantitative analysis of the fetal brain anatomy such fetal brain biometry~\cite{pier_3d_2016, kyriakopoulou_normative_2017, khawam_fetal_2021}.

Quantitative analysis of imaging biomarkers has defined cortical development for the typically developing brain \cite{rajagopalan_local_2011, clouchoux_normative_2012, wright_automatic_2014, xia_fetal_2019}, while other works evidenced discrepancies in cortical volumes and sulcal patterns in the pathological brain \cite{clouchoux_delayed_2013, egana-ugrinovic_differences_2013, im_quantification_2013, tarui_disorganized_2018}. 
Nevertheless, these analyses require prior additional segmentation processing steps. While expert manual image annotation is considered the gold standard, it is time-consuming, tedious, and subject to inter- and intra- expert variability. For this reason, manual segmentation is not an enduringly reliable method.

Thus robust automated fetal brain MRI segmentation methods are key for further analysis. Nonetheless, in contrast to adult brain segmentation, fetal brain segmentation remains challenging as to provide an age-invariant method~\cite{makropoulos_review_2018}.

\subsection{Related works in cortical plate segmentation}
\label{sec:related_works}
\begin{figure}[h!]
    \centering
    \includegraphics[width=0.8\linewidth]{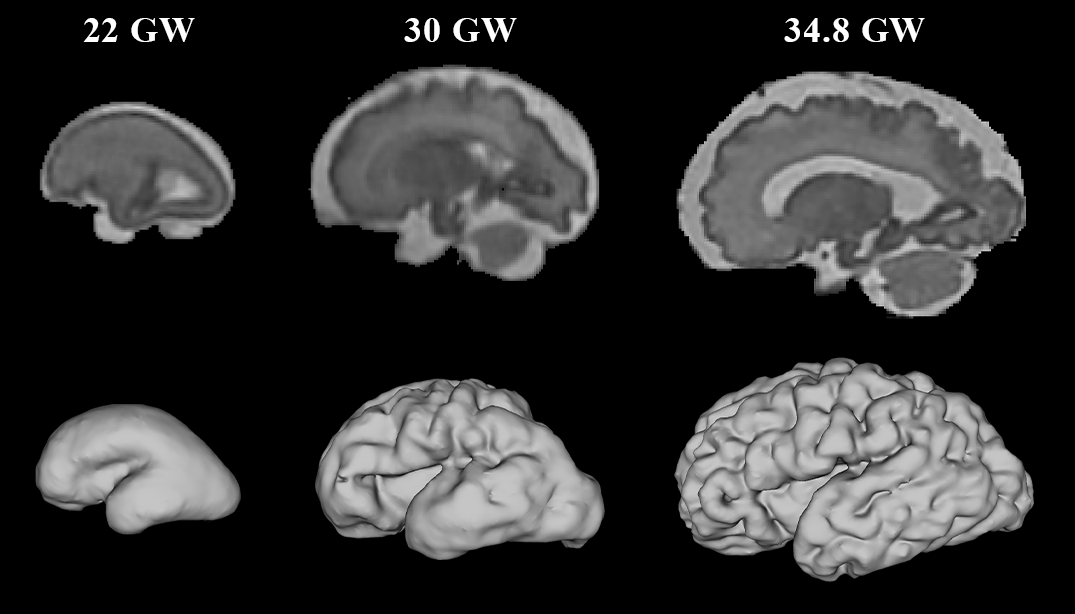}
    \caption{Illustration of MR T2w contrast and  \textit{in utero} cortical folding changes throughout gestation for neurotypical fetal brain subjects of 22, 30, and 34.8 gestational weeks (GW).}
    \label{fig:Fig_morpho_f_GA}
\end{figure}

Segmentation of the cortical plate is particularly challenging as it undergoes significant changes throughout gestation due to brain growth and maturation, respectively modifying the morphology and the image contrast (see Figure~\ref{fig:Fig_morpho_f_GA}). Furthermore, the cortex being a thin structure (from 1 to 2 mm for fetuses between 21 and 40 weeks of gestational age~\cite{vasung_quantitative_2016}) strongly affected by partial volume effects, anatomical topology is prone to be incorrectly captured by automatic segmentation methods. 

In Caldairou et \textit{al}.~\cite{caldairou_segmentation_2011}, they introduced the first topological-based segmentation of the fetal cortex based on geometrical constraints along with anatomical and topological priors. However, the sample size in that study was small (i.e. six fetuses) and topological correctness was not evaluated for the provided segmentations. More recently, deep learning (DL) methods have also focused on fetal brain MRI cortical gray matter segmentation. 
Based on a neonatal segmentation framework, a recent study introduced a hybrid segmentation process that minimizes the need for human interaction in the segmentation of the developing cortex \cite{fetit_deep_2020}. 
Also with DL models, a multiple-predictions approach with a test-time augmentation to improve the robustness of the method was suggested \cite{hong_fetal_2020}. 
Finally, a two-stage segmentation framework with an attention refinement was proposed \cite{dou_deep_2021}. 
Nevertheless, while the segmentation accuracy of these recent DL-based methods is promising, none of these CP segmentation works includes topological constraints nor assesses the topology. Overall, these studies of automatic segmentation frameworks report high similarity, as for overlap (Dice similarity coefficient of 0.87 in~\cite{dou_deep_2021} and 0.90 in~\cite{hong_fetal_2020}), and low difference, as for the boundary distance-based metrics (average symmetric surface distance of  0.28 $mm$ in~\cite{dou_deep_2021} and mean surface distance of 0.18 $mm$ in~\cite{hong_fetal_2020}), compared to the ground truth segmentation, but illustrated results show a lack of topological consistency with notably discontinuous/broken cortical ribbons (see Fig. 5 in ~\cite{dou_deep_2021}, Figures 5 and 6 in~\cite{fetit_deep_2020}). 

Here, we propose to integrate a topological constraint in a deep image segmentation framework to overcome the limitation of disjoint CP segmentation in fetal MRI. To our knowledge, only two works previously explored the topological fidelity of the semantic medical image segmentation with DL. 
First, Hu et \textit{al.}~\cite{hu_topology-preserving_2019} proposed using a topological loss for neuronal membrane segmentation. 
Second, a study presented topological constraints for MR cardiac image segmentation~\cite{clough_topological_2021} based on prior topological knowledge, such as the number of connected components or handles present in the structure of interest. Although theoretical CP topological features are known, such prior information could only be valid in a whole 3D volume analysis. Therefore, we inspired from~\cite{hu_topology-preserving_2019} that does not share this limitation to build a prior-free framework.

\subsection{Contributions}
\label{sec:contributions}
In this work, we incorporate topological constraints and assess the topology of the CP segmentation in fetal brain MRI. To this end, our first contribution is the generalization of a topological loss function based on persistent homology into a multi-dimensional (in topological spaces) formulation.
Using a publicly available dataset of SR reconstructed fetal brain MRI \cite{payette_automatic_2021} along with a spatiotemporal atlas \cite{gholipour_normative_2017}, we compare our topology-integrative optimization method (\textit{TopoCP}) to other widely used loss functions (\textit{Baseline} and \textit{Hybrid})~\cite{payette_fetal_2022} with a state-of-the-art U-Net segmentation network. We further assess our proposed method compared to semi-automatic manual annotations. 
For the first time, the topological correctness of the segmentation is assessed in the evaluation of automatic CP segmentation. In that respect, 
our second contribution is a new topology-based metric for the quantification of the CP segmentation defects not only in number but also in size.
We quantitatively compare automatic and semi-automatic segmentation with complementary metrics on two independent pure testing sets. As a complement, three fetal brain MRI experts further visually compare automatic segmentation on an additional out-of-distribution clinical dataset. 
The results evidence an overall significant improvement in the segmentation using our proposed topological loss function. 

In Section~\ref{sec:methodology} we introduce our multi-dimensional topological loss function, the overall segmentation framework and our original topology hole ratio assessment metrics; in Section~\ref{sec:experiment_design}, we describe the experimental design, along with materials and description of the experiments performed; in Section~\ref{sec:results_and_discussion}, we describe and discuss the results; and finally, we conclude on this work in Section~\ref{sec:conclusion}.

\section{Methodology}
\label{sec:methodology}
We propose a topologically-guided deep learning framework for the cortical plate segmentation of the fetal brain MRI. This is done by including a topological constraint in the optimization of state-of-the-art deep-learning image segmentation strategies. First, we will introduce our custom loss function that we adapted for generalization from~\cite{hu_topology-preserving_2019} (see Section~\ref{ssec:topological_loss}). Second, we present the segmentation framework in which our custom loss function is integrated (see Section~\ref{ssec:segmentation_framework}). Finally, we present topology-based metrics for further assessment of our method (see Section~\ref{ssec:topological_assessment}).

\subsection{Topological loss}
\label{ssec:topological_loss}
In semantic image segmentation, conventional optimization loss functions (e.g. the cross-entropy) often proceed to a pixel-wise comparison of the class-prediction that is summarized in a likelihood function $f$ to the one-hot encoded target vector. 
In this work, we aim to integrate the analysis of the global shape correctness of the prediction through the study of topology during the model optimization.

\subsubsection{Computational topology}
\label{sssec:computational_topology}
Topology defines the properties of an object that are preserved through deformation~\cite{boissonnat_computational_2006}.
In computational topology, local features are derived to conclude on the global properties of an object. Specifically, in a binary image, local information relies on the connectivity of a voxel to its neighbors in an objects. By generalizing the connectivity information, one can conclude on global features such as the number of connected components or holes.

\begin{figure}[h!]
    \centering
    \includegraphics[width=\linewidth]{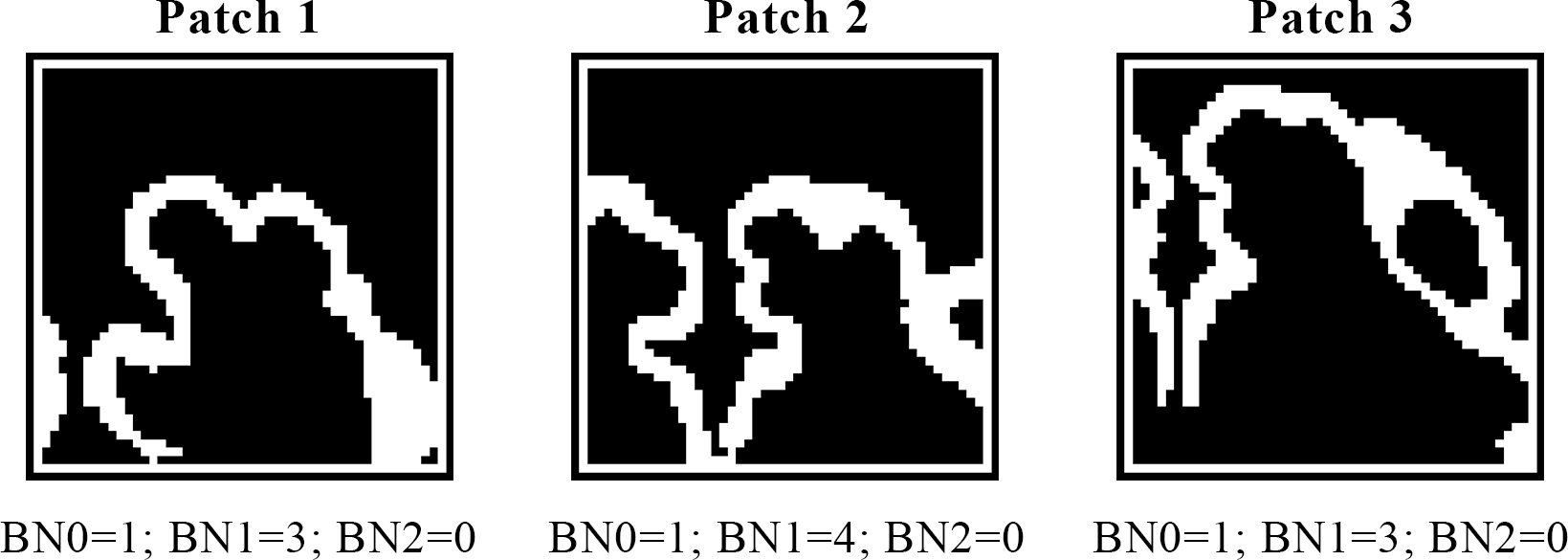} 
    \caption{Example patches of 2D binary images with the topological properties: number of connected components BN$0$, number of holes BN$1$, and number of cavities BN$2$.}
    \label{fig:Fig_Topology_binary_image}
\end{figure}

Topology structures are defined in the different topological dimensions. 
The number of topological structures in each $k$-dimension is counted with the $k$-dimensional Betti numbers (BN$k$).
In the context of 3D binary images, one can count BN$0$ connected components, BN$1$ holes (also called handles or tunnels), and finally BN$2$ cavities. 
Figure~\ref{fig:Fig_Topology_binary_image} shows three 2D binary image patchs of the cortex and their Betti numbers. Note that for a 2D binary image, BN$2$ is always 0.

In practice, and similarly to~\cite{hu_topology-preserving_2019}, prior to the computation of topology, we pad the image patches twice with constant values (see Figure~\ref{fig:Fig_Topology_binary_image}, Figure~\ref{fig:Fig_persistent_homology}~(A) and Figure~\ref{fig:Fig_loss_topo}~(A)). We perform a first padding with the maximal value of the patch in order to work with closed structures. Then, we pad the patch with 0 value voxels to define a background. 

\subsubsection{Persistent homology}
\label{sssec:persistent_homology}

\begin{figure}[h!]
    \centering
    \includegraphics[width=\linewidth]{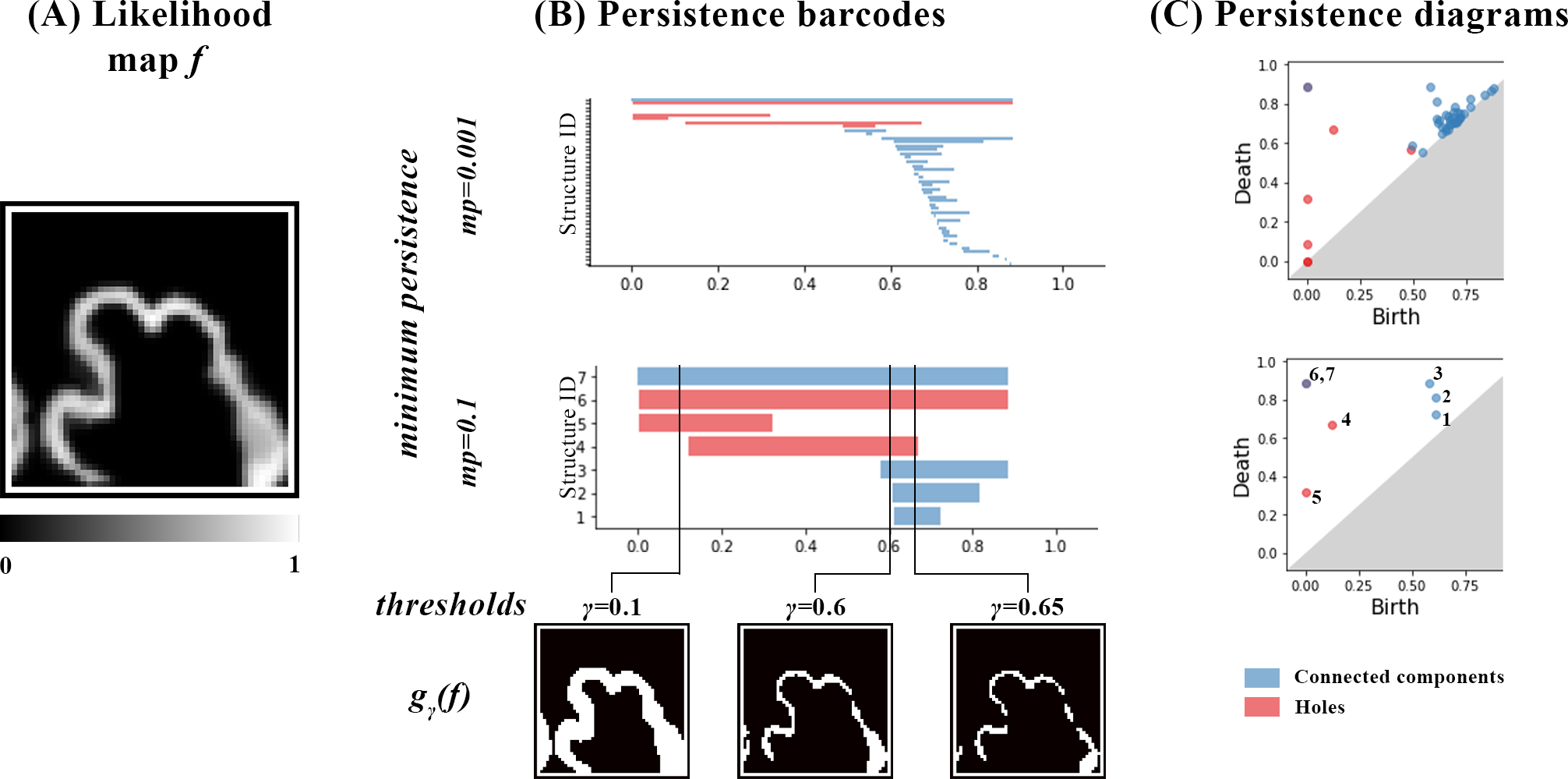} 
    \caption{Illustration of persistent homology for a CP likelihood map $f$ (A). Panel (B) illustrates the progressive filtration of $f$ with minimum persistence (mp) of 0.001 (top) and 0.1 (bottom). Topological structures are reported into a persistence barcode. Panel (C) shows the persistence diagram for both mp. 
    In the persistence barcode, respectively diagram, blue bars, respectively blue dots, represent the connected components. Similarly, red elements represent the 1-dimensional holes.}
    \label{fig:Fig_persistent_homology}
\end{figure}

Persistent homology offers a workaround to analyze the topology of a continuous-valued  $n$-dimensional image function. 
In the context of our semantic image segmentation, we consider the likelihood map  $f \colon \Omega \subset \mathbb{R}^{n} \rightarrow [0,1]$ of a voxel to belong to the CP that is predicted from a DL-based model (Figure~\ref{fig:Fig_persistent_homology} (A)).
In order to reduce the problem to a binary analysis as presented in the previous Section~\ref{sssec:computational_topology}, persistent homology tracks the topological structures of $f$ through filtration $g_\gamma$ to different thresholds $\gamma \in [0,1]$:

\begin{align}
  g_{\gamma} \colon [0,1] &\longrightarrow \{0,1\} \notag \\
  x &\longmapsto g_{\gamma}(x) =
        \begin{cases} 
        1 & \text{if } x \geq \gamma, \\
        0 & \text{otherwise.}
\end{cases}
\end{align}

Snapshots of the topology are reported into a persistence barcode (Figure~\ref{fig:Fig_persistent_homology} (B)). Each bar corresponds to a topological structure (e.g. connected components, handles) which is characterized by its appearance and disappearance threshold values $(\gamma_{birth}; \gamma_{death})$. 
The persistence barcode can be filtered based on the structures persistence. The persistence of a structure is defined by its life time $\Delta\gamma = \gamma_{death}-\gamma_{birth}$. In persistent homology, the minimum persistence (mp) is the minimum lifetime accepted in the topological structures filtration. 
Figure~\ref{fig:Fig_persistent_homology} (B) shows an example of a likelihood image binarized at different thresholds $\gamma \in [0,1]$. Two persistence barcodes with $mp=0.001$ (top) and $mp=0.1$ (bottom) are presented. With a lower mp, we observe the presence of a considerable amount of irrelevant structures. 
Indeed, the CP is a thin cerebral tissue (only a few voxels-wide in SR volumes), and is therefore sensitive to broken connections. In the barcode, this turns into the appearance of many connected components (blue) with short life time, i.e. low persistence. 

Finally, the structure pairs $(\gamma_{birth}; \gamma_{death})$ extracted from the persistence barcodes are considered as coordinates. These coordinates define critical points, transcribed in the associated persistence diagram (see Figure~\ref{fig:Fig_persistent_homology}~(C)). With low mp (top) we observe many critical points close to the diagonal. This diagonal corresponds to $\gamma_{birth}=\gamma_{death}$, i.e. the structure does not exist.

The choice of the mp must be set to avoid noise structures without being too strict. Note that for a binary image, all topological structures have a persistence $\Delta\gamma=1$, with coordinates $(0;1)$ in the persistent diagram representation.

\subsubsection{Topological loss function for fetal brain MRI}
\label{sssec:loss_function}
The topological loss function aims at directly comparing the persistent homology of the predicted likelihood map $f$ to the target true topology. We propose a topological loss function $\mathcal{L}_{topo}$ that is adapted from~\cite{hu_topology-preserving_2019}. Our contribution lies in the multi-dimensional approach of the topological loss computation. 
While our focus in this work is on the 1-dimensional holes, we evidence in Section~\ref{sssec:persistent_homology} the importance the connected components can have in the persistent homology of the fetal CP. Therefore, differently from  \cite{hu_topology-preserving_2019} that only considered $1$-dimensional structures (i.e. 2D holes), our topological loss for the fetal CP segmentation will additionally integrate the analysis of $0$-dimensional homology structures. 

\begin{figure}[h!]
    \centering
    \includegraphics[width=\linewidth]{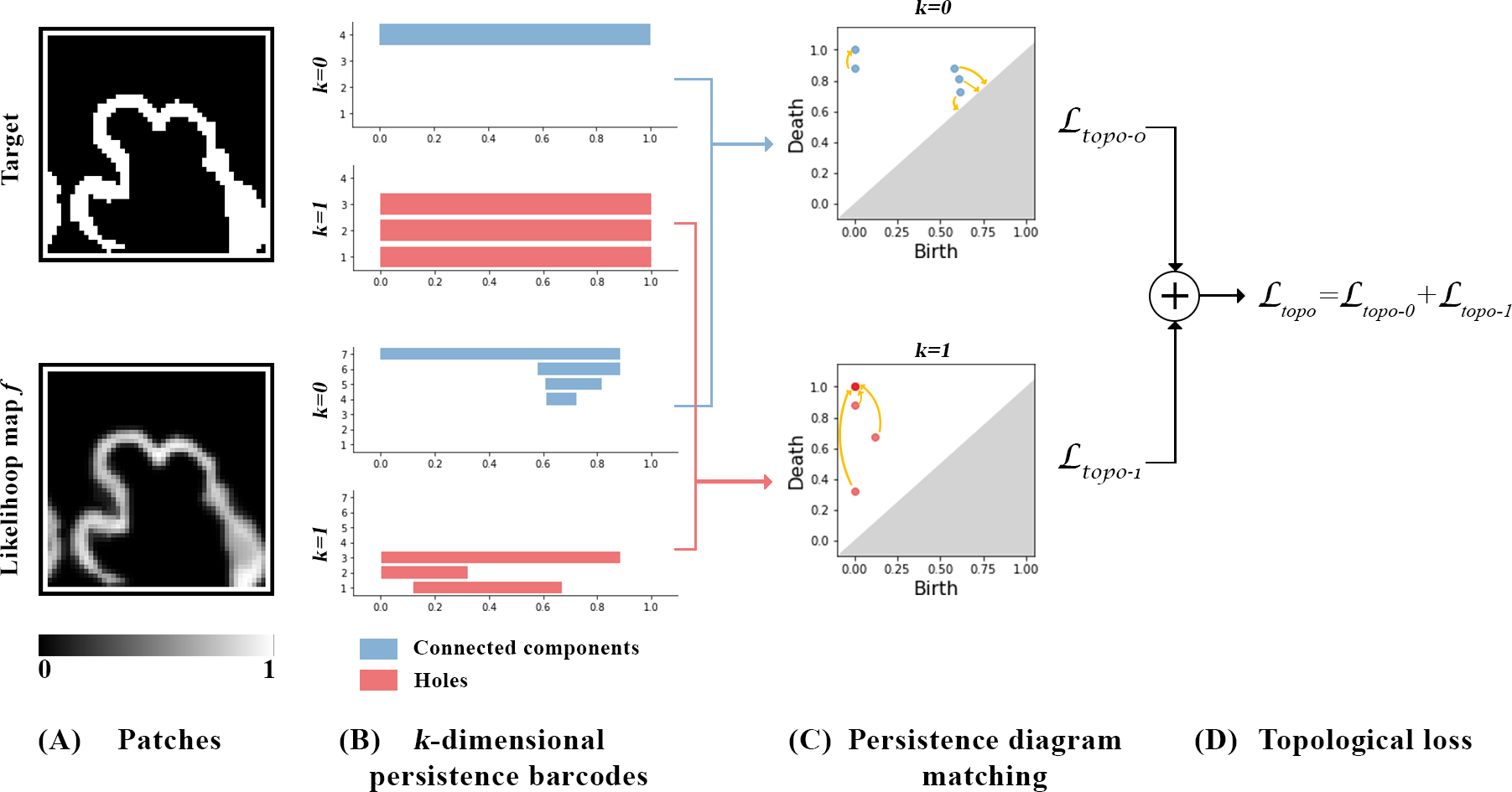} 
    \caption{Illustration of the topological loss computation process between a ground truth binary image and a likelihood map $f$ (A). 
    (B) shows their corresponding persistence barcodes in each dimension ($k=0$ and $k=1$). 
    (C) illustrates a per-dimension persistence diagram matching. 
    (D) shows how the final $\mathcal{L}_{topo}$ is inferred, with $\omega_{0}=\omega_{1}=1$. 
    In the persistence barcode, respectively diagram, blue bars, respectively blue dots, represent the connected components. Similarly, red elements represent the holes. }
    \label{fig:Fig_loss_topo}
\end{figure}

We detail here the process of the computation of $\mathcal{L}_{topo}$ between two 2D image patches, the target segmentation and the predicted likelihood map $f$ (see Figure~\ref{fig:Fig_loss_topo} (A)), hence the dimensions involved are $k \in \{0;1\}$. 
First, persistent homology is computed tracking all $k$-dimensional structures in both images (see in Figure~\ref{fig:Fig_loss_topo} (B), the persistence barcodes partitioned by dimension). 
Second, the per-dimension persistence diagrams are matched between the ground truth and the prediction (Figure~\ref{fig:Fig_loss_topo} (C)).
All $k$-dimensional structures are matched such that, the $N$-greater persistence structures are matched to the $N$ true ground truth elements. Note that a sufficiently accurate likelihood is needed to prevent structures mismatch. Others are matched to the diagonal.
Based on the implementation in~\cite{hu_topology-preserving_2019} ~\footnote{https://github.com/HuXiaoling/TopoLoss}, we compute an adapted Wasserstein distance~\cite{cohen-steiner_lipschitz_2010}, from the matched pairs of critical points in each dimension. 
The $k$-dimensional distance is our $\mathcal{L}_{topo-k}$, the $k$-dimensional topological loss function.
Ultimately, $\mathcal{L}_{topo-k}$ are combined such that : 

\begin{equation}
\mathcal{L}_{topo}=\sum_{k=0}^{K}\omega_{k}\mathcal{L}_{topo-k}
\label{eq:topo_loss}
\end{equation} 

where $\mathcal{L}_{topo-k}$ is the topological loss of the $k$-dimensional space with a contribution weight of $\omega_{k}$.

\subsection{Segmentation framework}
\label{ssec:segmentation_framework}
Our topological loss is an architecture-agnostic optimization function. In other words, it is independent of the deep-learning network architecture. In our segmentation framework, we use a state-of-the-art architecture, U-Net~\cite{navab_u-net_2015},  to compare different optimization methods. Two reference loss functions (\textit{Baseline} and \textit{Hybrid}) are implemented to evaluate the added value of our topological loss function (\textit{TopoCP}) (see configurations in Section~\ref{sssec:training_strategy}). 

\subsubsection{Model architecture}
\label{sssec:models_architecture}
The well established U-Net~\cite{navab_u-net_2015} architecture is selected as it has recently proved its good accuracy in fetal brain MRI tissue segmentation~\cite{khalili_automatic_2019, payette_automatic_2021, payette_fetal_2022}.
We use a 2D U-Net architecture that is composed of an encoding and a decoding paths with skipped connections.
The encoding path in our study is composed of 5 repetitions of the followings: two 3x3 convolutional layers, followed by a rectified linear unit (ReLU) activation function and a 2x2 max-pooling downsampling layer. Feature maps are hence doubled after each block, starting from 32 to 512. In the expanding path, 2x2 upsampled encoded features concatenated with the corresponding encoding path are 3x3 convolved  and passed through ReLU. The network prediction is computed with a final 1x1 convolution.
The number of network trainable parameters is  7,852,002.

\subsubsection{Multiview patch-based approach}
\label{sssec:preprocessing}
First, T2w images are masked 
in order to only consider intracranial space voxels in the CP segmentation.
Second, to alleviate the computational cost due to the topological loss, $64 \times 64$ voxel size sub-image patches are extracted from the 3D volume in the three orthogonal planes (axial, coronal and sagittal). 
Bringing information from the three dimensional orientations, our method thus implements a 2.5D, or multiview, patch-based approach.
To increase the number of predictions per voxel, overlapping patches are extracted. Empirically, the sliding window's step size for the patch extraction is set 16 voxels. 
Finally, intensities of all patches are standardized to have mean 0 and variance 1. 

\subsubsection{Training and optimization strategies}
\label{sssec:training_strategy}

Input samples are randomly augmented at each epoch of the training phase. Extensive augmentations are performed spatially (flipping and elastic deformation) and intensity-based (bias, blurring, gamma and noise). All augmentation have a probability of occurrence of $0.5$, except flipping occuring with a probability of $0.2$.
Augmentation are performed with the TorchIO python package~(v0.18.75)~\cite{perez-garcia_torchio_2021}. 

Two reference segmentation methods (\textit{Baseline} and \textit{Hybrid}) are trained with fetal brain MRI state-of-the-art optimization loss function, in order to compare with our method (\textit{TopoCP}). Thus, we evaluate the following three configurations:
\begin{itemize} 
    \item \textit{Baseline} is trained using the distribution-based binary cross-entropy loss function $\mathcal{L}_{bce}$. 
    
    \item \textit{Hybrid} is trained with an hybrid loss function combining the dice loss $\mathcal{L}_{dice}$ and the binary cross-entropy loss $\mathcal{L}_{bce}$ such that: 
    \begin{equation}\mathcal{L}=\mathcal{L}_{bce}+\mathcal{L}_{dice}\end{equation}
    Such hybrid loss function proved efficient in multi-tissue fetal brain MRI segmentation, as it has been used by the Top 5 teams of the 2022 edition of the MICCAI FeTA challenge~\cite{payette_fetal_2022}.
    
    \item \textit{TopoCP} is trained with the following loss combination: 
    \begin{equation}\mathcal{L}=(1-\lambda_{topo})\mathcal{L}_{bce}+\lambda_{topo}\mathcal{L}_{topo}\end{equation} 
    where $\mathcal{L}_{bce}$ is the binary cross-entropy loss and $\mathcal{L}_{topo}$ is the topological term presented in Section \ref{sssec:loss_function}. $\lambda_{topo}$ defines the weight of the contribution of \begin{math}\mathcal{L}_{topo}\end{math} in the final loss.
\end{itemize}

As the computation of the topological loss is expensive and need sufficiently accurate probability maps to perform a relevant matching of the structures (see Section~\ref{sssec:loss_function}), we adopt the training strategy presented in \cite{hu_topology-preserving_2019}: 
1) a common warm-up network is trained over 15 epochs using the binary cross-entropy loss $\mathcal{L}=\mathcal{L}_{bce}$; 
2) \textit{Baseline}, \textit{Hybrid} and \textit{TopoCP} are initialized with the pretrained \textit{warm-up} weights. An early stopping strategy monitors the global validation loss $\mathcal{L}$, respectively the topological validation loss $\mathcal{L}_{topo}$, for the \textit{Baseline} and \textit{Hybrid} configurations, respectively for our \textit{TopoCP} configuration. All learning rates are set to $0.01$. 
Training and evaluation were performed with Tensorflow v2.7 ~\cite{martin_abadi_tensorflow_2015} and a GeForce RTX 2080TI GPU.

A 4-folds cross-validation approach is adopted to assess the learning performances of the different methods. In this way, we will assess multiple $\lambda_{topo}$ in order to determine an optimal value (see Section~\ref{sssec:lambda_topo_settings}).

\subsubsection{Ensemble learning}
\label{sssec:postprocessing}
In order to reduce the variance and increase the generalization power of our model, we adopt \textit{ensemble learning}. In the final testing inference phase of each configuration, we perform a majority voting on the summed likelihoods from all 4 cross-validation networks. Finally, despite cortex would theoretically be two connected components (left and right hemispheres), in practice, partial volume in the mid-sagittal plane most often leads to having one single component. Thus, only the biggest connected component of the \textbf{whole cortical volume }ensemble prediction is kept.

\subsubsection{\textit{TopoCP} parameters setting} 
\label{sssec:parameters_settings}

As introduced in Section~\ref{sssec:persistent_homology}, persistent homology structures are filtered on a minimum persistence criteria. 
This criteria tunes the sensitivity of our loss to the noisy structures out of the filtration step.
Empirically, we observed that the higher minimum persistence is the tougher filtration of the structures and thus may discard relevant ones. Reducing the persistence threshold leads to having an increasingly large formation of noisy irrelevant structures to be matched. Based on these empirical analysis, we set our minimum persistence to $0.01$ for all experiments. 

Equation~\ref{eq:topo_loss} presents our global topological loss in which different contributions can be assigned to each dimension. 
We analyzed the importance of both 0-dimensional and 1-dimensional topological terms on a reduced set of randomly sampled patches.
Empirically, we observe that the importance of 0-dimensional and 1-dimensional topological terms is patch-dependent. In some patches, $\mathcal{L}_{topo-0}$ is more affected than $\mathcal{L}_{topo-1}$, and vice versa in others. 
We therefore decided to give equal contribution to both terms, as they are undoubtedly both important. In other words, all $k$ dimension had the same weighting \textbf{$\omega_{k}=1$}.

Finally, the contribution of our global topological loss is valued with the \textbf{$\lambda_{topo}$}. We describe in Section~\ref{sssec:lambda_topo_settings} the cross-validation approach implemented to determine an optimal value.

\subsection{Topological assessment of the fetal CP}
\label{ssec:topological_assessment}
Recent works~\cite{yeghiazaryan_family_2018,maier-hein_metrics_2022} evidenced the importance of considering complementary metrics for the assessment of semantic segmentation methods. Specifically, in this work, segmentation should be assessed for its closeness to the target topology.

We presented in Section~\ref{sssec:computational_topology} the BN$k$ that define global features of a 3D binary image. To quantitatively compare topology of binary images, we introduce the $k$-dimensional Betti number error (BNE$k$) as the absolute difference of the ground truth expected value and the prediction measures.
As the CP segmentation is filtered for its biggest connected component (see Section~\ref{sssec:postprocessing}), BNE$0$ is incidentally irrelevant to consider.
Additionally, this work specifically focuses on the presence of 1-dimensional holes. In this way, we only consider BNE$1$. 
Besides, considering that the human cortex is a closed structure with no obstruction, its ground truth expected BN$1$ is 0. 

While BNE$1$ focuses on the count of occlusions on the CP surface, this score is not providing any information on the holes themselves. Indeed, we often observe that the number of holes that is a discrete value is not necessarily correlated to the size of the broken connections (i.e. the obstruction) nor to the size of the structure of interest (i.e. the fetal CP). 

In that respect, we introduce a new metric that aims to quantify the size of the hole. The hole ratio (HR) is the ratio of false negative voxels that are connected to a hole ($FN_{holes}$) over the true voxels of the region of interest, which are represented by the true positives ($TP$) and false negatives ($FN$).

\begin{equation}
\label{eq:hole_ratio}
    HR = \frac{FN_{holes}}{TP + FN}
\end{equation}

\begin{figure}[h!]
    \centering
    \includegraphics[width=\linewidth]{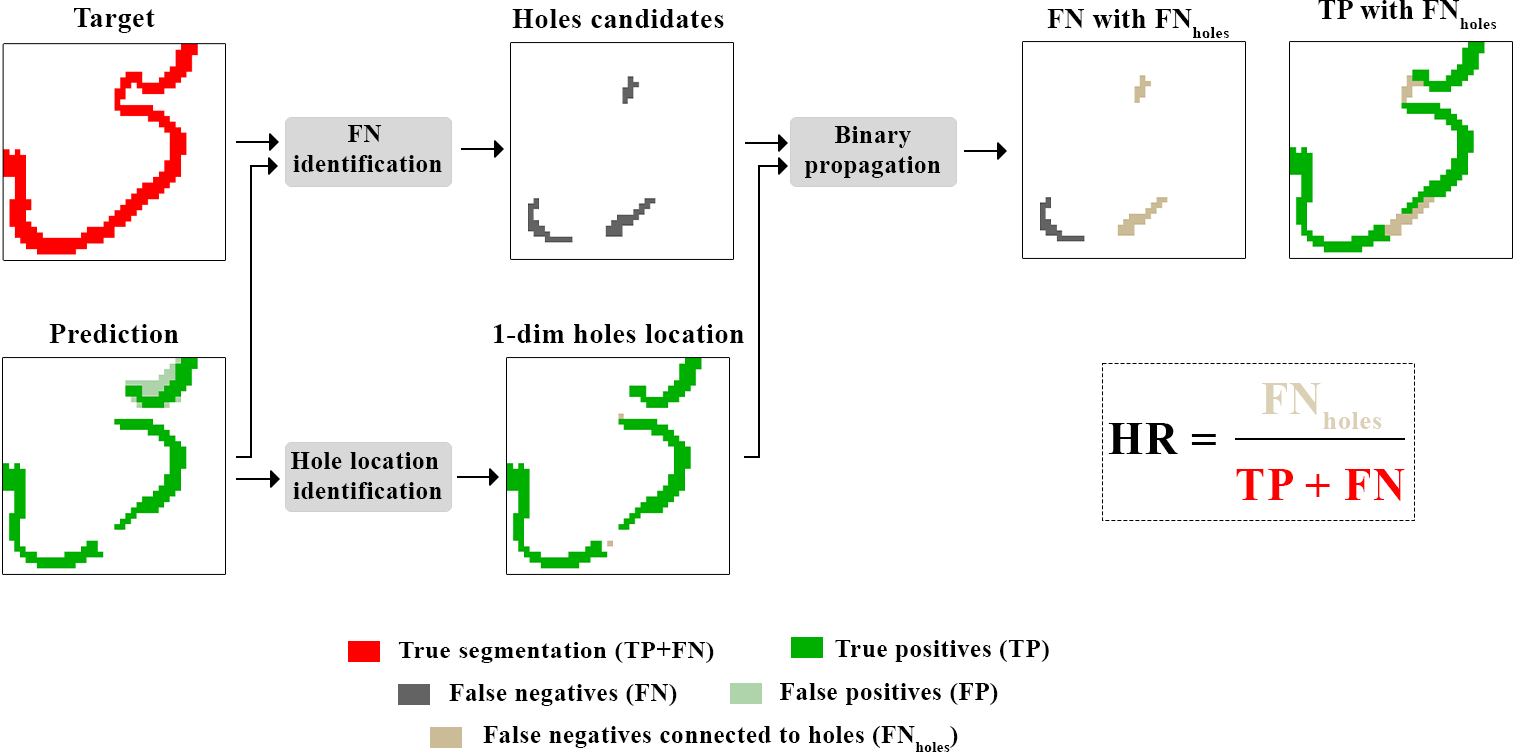} 
    \caption{Illustration of the workflow for the computation of the Hole Ratio (HR). }
    \label{fig:Fig_hole_ratio}
\end{figure}

Our in-house implementation (available online, see Section~\ref{ssec:code_availability}) illustrated in Figure~\ref{fig:Fig_hole_ratio} successively identifies the location of one voxel per 1-dimensional hole, propagates these voxels into the mask of holes candidates, i.e. the FN, and finally, computes the volume ratio presented in Equation~\ref{eq:hole_ratio}. Let us note that this measure strongly relies on the topological correctness of the ground truth.  

\begin{figure}[htb]

    \centering

\begin{subfigure}{0.8\linewidth}
\centering
    \includegraphics[width=\linewidth]{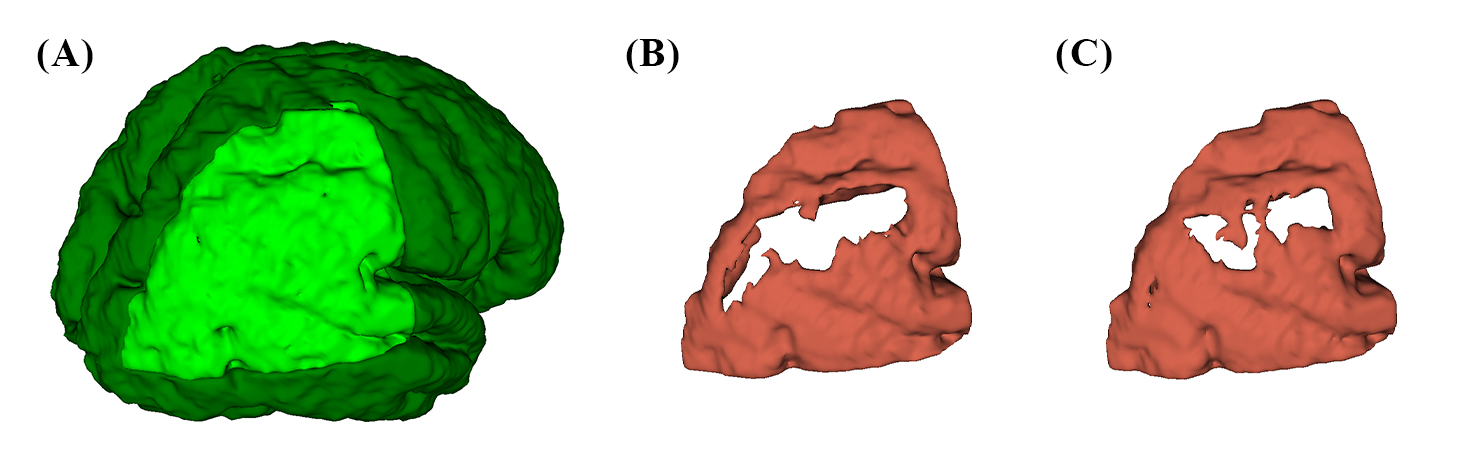} 
    \caption{3D rendering of (A) Ground truth subject cortical plate (dark green) and region of interest (light green), and two different segmentation (B) and (C).}
    \label{fig:Fig_N_trous_vs_size}
\end{subfigure}

\medskip

\begin{subfigure}{0.8\linewidth}
    \centering
\begin{tabular}{c | c  c  c}
 &  (B) & (C) \\
\hline
BNE1     & 1 & 4    \\
HR      & 0.36 & 0.33   \\
\end{tabular}
\caption{\label{tab:metrics_bn1_hr}
Table of the topology-based metrics (BNE1: 1-dimensional Betti number error; HR: Hole ratio) for the example cortical plate segmentation (B) and (C) shown in Figure~\ref{fig:Fig_N_trous_vs_size}.}
\end{subfigure}
\label{fig:composite_figure_top_bottom}

\caption{Illustration of the discrepancy between the quantification of holes as a number and its quantification as a ratio over the region of interest.}
    
\end{figure}

Figure~\ref{fig:Fig_N_trous_vs_size} shows (B) and (C), two 3D rendering of the same image patch CP segmentation. (A) shows the 3D rendering of the ground truth, enlighting the region of interest. One can easily observe that the (B) segmentation has a main hole compared to the (C) segmentation that presents multiple medium-size holes. Additionally, quantitative results (see Table~\ref{tab:metrics_bn1_hr}) confirm the discrepancy between the quantity of holes as a number and the quantity of holes as a ratio of the region of interest. 

Albeit BNE1 must be used with caution, it is still a relevant score to assess the CP topology in the absence of topologically accurate ground truth segmentation. Nonetheless, we promote the use of an additional quantitative metric relative to the size of the occlusions to undertake a robust quantitative analysis of the broken connections in the CP segmentation.

Both topology-based measures (BNE1 and HR) rely on the cubical complex implementation of the GUDHI library (v3.5.0)~\cite{noauthor_gudhi_nodate}.

\subsection{Code availability} 
\label{ssec:code_availability}

\textit{Baseline}, \textit{Hybrid} and \textit{TopoCP} models implementation, including the optimization loss functions can be found in our Github repository~\footnote{\url{https://github.com/Medical-Image-Analysis-Laboratory/FetalCP_segmentation}}. The weights of the trained model are made available to perform inference. The implementation of the topology-based evaluation metric is also made available at this link.

\section{Experiment design}
\label{sec:experiment_design}

\begin{figure}[h!]
    \centering
    \includegraphics[width=\linewidth]{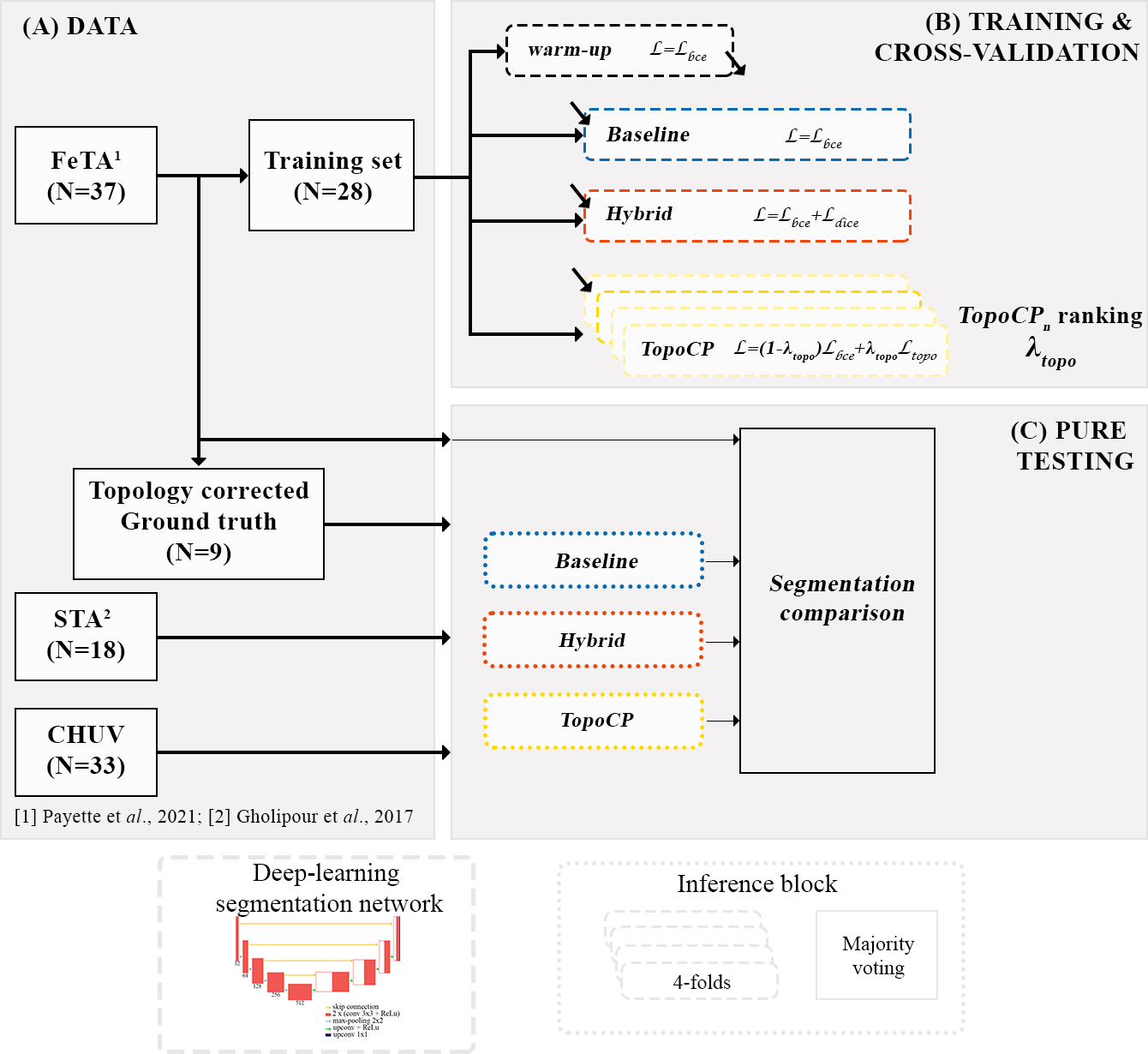} 
    \caption{Illustration of the overall experiment design. Panel (A) illustrates the different datasets and their splitting for training/testing purposes (see Section~\ref{ssec:datasets}). Panel (B) illustrates the training phase. A common warm-up network is trained to initialize the three configurations \textit{Baseline}, \textit{Hybrid} and \textit{TopoCP}, each optimized with its own optimization loss function (see Section ~\ref{ssec:segmentation_framework}). A cross-validation approach is used to determine an optimal hyperparameter $\lambda_{topo}$ (see Section~\ref{sssec:lambda_topo_settings}). Panel (C) illustrates the testing phase. Predictions are inferred through the cross-validation networks and combined in a majority voting step. Methods are assessed and compared quantitatively  with complementary performance metrics (see Sections~\ref{sssec:method_comparison},~\ref{sssec:ga_analysis}~\ref{sssec:spatial_analysis}, ~\ref{sssec:group_analysis}~\ref{sssec:weak_labels_analysis}) and qualitatively by three experts (see Sections~\ref{sssec:experts_evaluation}).}
    \label{fig:Fig_experiment_design}
\end{figure}

The overall experiment design to compare the three segmentation frameworks \textit{Baseline}, \textit{Hybrid} and our \textit{TopoCP} is outlined in Figure~\ref{fig:Fig_experiment_design}.

\subsection{Datasets}
\label{ssec:datasets}

A summary of clinical and atlas datasets is shown in Table~\ref{tab:data_summary}.

\begin{table*}[h!]
    \centering
    \resizebox{\textwidth}{!}{%
\begin{tabular}{c   c   c c c c  }
\toprule

 & Dataset  & \thead{Number\\of subjects\\(Neurotypical\\/\\Pathological)} & \thead{Gestational\\age (weeks)} & \thead{SR reconstruction} & \thead{Image \\resolution\\(mm$^{3}$)} \\ 
\midrule

TRAINING  & \thead{FeTA\\\cite{payette_automatic_2021}} & \thead{24\\(13/11)} & \thead{[20.9-34.8]\\ (28.2$\pm$3.6)} & \thead{Simple-IRTK\\~\cite{kuklisova-murgasova_reconstruction_2012}} & $0.86^{3}$\\ 
\cmidrule(lr){1-1}\cmidrule(lr){2-6}

\parbox[t]{3mm}{\multirow{6}{*}{\rotatebox[origin=c]{90}{\parbox[c]{2cm}{\centering TESTING }}}} & \thead{FeTA\\\cite{payette_automatic_2021}} & \thead{9\\(4/5)} & \thead{[22.9-34.8]\\ (27.4$\pm$3.6)} &  \thead{Simple-IRTK\\~\cite{kuklisova-murgasova_reconstruction_2012}} & $0.86^{3}$\\ 
 & \thead{STA\\\cite{gholipour_normative_2017}} & \thead{18\\(18/0)} & [21-38] & \thead{Gholipour et. al, 2017 \\\cite{gholipour_normative_2017}} & $0.80^{3}$\\  
 & \thead{CHUV} & \thead{33\\(24/9)} & \thead{[21-35]\\ (29.6$\pm$3.6)}  & \thead{MIALSRTK \\\cite{tourbier_sebastien_sebastientourbiermialsuperresolutiontoolkit_2019}} & $0.80^{3}$\\ 

\bottomrule

\end{tabular}}
\caption{Summary of the data used for training and quantitative and qualitative evaluation.}
\label{tab:data_summary}
\end{table*}

\subsubsection{Clinical dataset: FeTA}
\label{sssec:clinical_dataset}

We use the subset of the publicly available dataset Fetal Tissue Annotation and Segmentation Dataset (FeTA v2.0)~\cite{payette_automatic_2021} with Simple-IRTK~\cite{kuklisova-murgasova_reconstruction_2012} SR-reconstructions at isotropic resolution of $0.86 mm$. After visual inspection of the images, seven volumes were excluded due to bad SR quality (3) and severe pathology (7) (e.g. major ventriculomegaly). The remaining 33 fetal brains were composed of 17 neurotypical and 16 pathological subjects, in the gestational age (GA) range of 20.9 to 34.8 weeks. 
Twenty-four (24) subjects (13 neurotypical and 11 pathological subjects in the GA range of 20.9 to 34.8 weeks, 28.2$\pm$3.6) were randomly selected for the method development and the remaining nine (9) subjects (4 neurotypical and 5 pathological subjects in the GA range of 22.9 to 34.8 weeks, 27.4$\pm$3.6) were retained for pure testing purposes.
Note that details on the fetal brain pathologies are not disclosed in the dataset information.

\begin{figure}[h!]
    \centering
    \includegraphics[width=\linewidth]{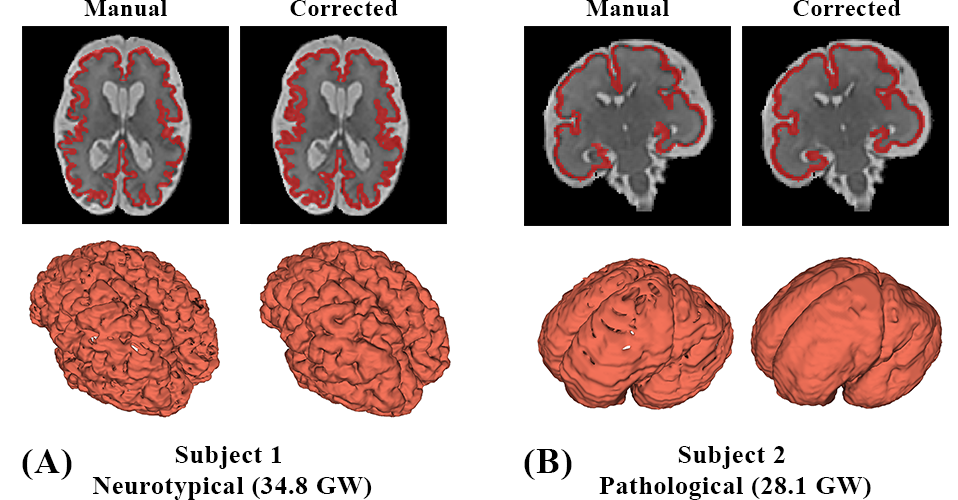} 
    \caption{Illustration of the FeTA original manual and the corrected ground truth CP segmentation for (A) Subject 1, a 34.8 weeks of GA neurotypical subject and (B) Subject 2, a 28.1 weeks of GA pathological subject. T2w images and CP segmentation overlaid (top) are respectively shown on an axial and coronal view for Subject 1 and Subject 2. 3D rendering is presented for all segmentations (bottom).}
    \label{fig:Fig_CP_segmentation_manual_refinement}
\end{figure}

Manual label annotations of the intracranial space tissues classified into seven categories (extra-axial cerebrospinal fluid spaces, the cortical gray matter (GM), the white matter, the ventricular system (lateral, third and fourth ventricles), the cerebellum, the deep gray matter and the brainstem) are provided for all SR reconstructed volume. In this work, we exclusively consider the cortical GM label.
Annotations were manually performed following an optimized protocol.
Two experts respectively annotated the external border of the cortex cerebri and the external border of the white matter, on every 2$^{nd}$ to 3$^{rd}$ slice of the axial view. 
Individual structure annotations are post-processed with interpolation and smoothing prior to merging into a final label maps.
Ultimately, sparse interpolated annotations result in noisy label maps often showing topological inconsistencies.
Figure~\ref{fig:Fig_CP_segmentation_manual_refinement} shows the extracted cortical GM from the final label maps (left) for (A) Subject~1, a 34.8 weeks of GA neurotypical subject and (B) Subject~2, a 28.1 weeks of GA pathological subject. Three-dimensional (3D) rendering evidences the presence of apertures in the final CP annotations. 

As motivated in Section~\ref{ssec:topological_assessment}, topologically accurate ground truth segmentation are necessary to perform a valid topological assessment of an automatic method. In this respect, we perform further manual correction of the CP FeTA manual annotations.
Four engineers refined the CP label maps of the 9 fetal brains of the clinical pure testing set. Editing of the label maps were performed using the ITK-SNAP~\cite{yushkevich_user-guided_2006} software with a specific focus on the topological correctness and contour refinement of the label maps.
Finally, all CP manual corrections were checked (and corrected if needed) by a pediatric radiologist with 17 years of experience. Right columns of panel (A) and (B) of Figure~\ref{fig:Fig_CP_segmentation_manual_refinement} show the corrected tissue annotations overlaid to the T2w image and their 3D rendering. In our further experiments, we refer to the corrected manual annotations as the ground truth.

The original FeTA dataset is under the ethical committee of the Canton of Zurich, Switzerland (Decision numbers: 2017-00885, 2016-01019, 2017-00167).

\subsubsection{Atlas dataset: STA}
\label{sssec:atlas_dataset}

The normative spatio-temporal atlas (STA) of the fetal brain~\cite{gholipour_normative_2017} provides 3D high-quality isotropic volumes for all gestational age between 21 and 38 weeks.
Each atlas subject is constructed with the contribution of 6 to 23 SR-reconstructed individual fetal brains. The integration of multiple subjects per gestational age reduces the morphological variability. Therefore, the T2w atlas images appear smoother than clinical acquisitions.

The atlas comes with two label maps, respectively containing the cerebral tissue and structure labels, and a regional cortex parcellation. 
The initial tissue label maps of more than 50 classes are converted into a 7-tissue label maps, matching those defined in the FeTA dataset. Minor adjustments are performed while synchronizing tissue and regional maps. For instance, voxel labelled as cortex in one map and as corpus callosum in the other are dumped in white matter class. As opposed to the clinical dataset, atlas labels do not require further manual corrections as they were already manually refined and present decent topology~\cite{gholipour_normative_2017}.

This dataset is approved by the Boston Children’s Hospital Institutional Review Board and the Committee on Clinical Investigation and written informed consent was obtained from all participants.

\subsubsection{Out-of-domain clinical dataset}
\label{sssec:clinical_dataset_chuv}

Thirty-three fetal brain clinical MR examination conducted in our institution at the Lausanne University Hospital (CHUV), Lausanne, Switzerland, were SR-reconstructed with the MIALSRTK pipeline~\cite{tourbier_efficient_2015,tourbier_sebastien_sebastientourbiermialsuperresolutiontoolkit_2019}. SR volumes are further resampled to an isotropic resolution of $0.8 mm$ and an engineer coarsely realigned the volumes to the anatomical plane. This clinical set is composed of 24 neurotypical and 9 pathological subjects in the GA range of 21 to 35 weeks (29.6 $\pm$3.6). 
The quality of the SR reconstructions was assessed similarly as in~\cite{khawam_fetal_2021} into three categories: bad (non usable, very blurred), average (overall good with remaining partial volume effect/blurring), and excellent (good quality with no blurring). Overall, none of the clinical SR reconstruction is bad, 17 are average 14 are excellent. No reference segmentation are available for this dataset.

The local ethics committee of the Canton of Vaud, Switzerland (CER-VD 2021-00124) approved the retrospective collection and analysis of MRI data and the prospective studies for the collection and analysis of the MRI data in presence of a signed form of either general or specific consent.

\subsection{Assessment metrics}
\label{ssec:assessment_metrics}
Although our segmentation framework infers CP segmentation in a 2.5D multi-view strategy (2D image patches from the three orthogonal planes), we proceed to the quantitative evaluation in 3D that is of the whole cortical volume. Automatic medical image segmentation requires the use of complementary metrics for the assessment of different aspects of the segmentation~\cite{yeghiazaryan_family_2018,maier-hein_metrics_2022}.
Most commonly, overlap-based (e.g. Dice similarity coefficient, Jaccard similarity coefficient, Intersection over union) and distance-based (e.g. X$^{th}$ percentile Hausdorff distance, average symmetric surface distance) metrics are reported~\cite{yeghiazaryan_family_2018}. However, in this work we aim to assess the segmentation not only in terms of overlap and distance accuracy to the ground truth but also in terms of shape correctness. Therefore, we also consider topology-based metrics. Table~\ref{table:metrics_summary} summarizes the metrics used in the training (for learning monitoring) and testing (for final evaluation) phases. 

The Dice Similarity Coefficient (DSC)~\cite{dice_measures_1945} is an overlap-based similarity metric. Robust to outliers, it is a widely-used metric to assess medical image segmentation accuracy. 
Average Symmetric Surface Distance (ASSD) is the mean of the directed average surface distances~\cite{yeghiazaryan_family_2018}. The latter is defined as the average of the distances of points from one surface to their closest points on the other object boundary. The ASSD is computed using the python MedPy~\footnote{https://loli.github.io/medpy/} implementation (v0.4.0).

In the case of absence of topologically accurate ground truth (i.e. see Section~\ref{sssec:lambda_topo_settings} for cross-validation details), we consider BNE1 to quantitatively assess the topology, while our proposed hole ratio HR is used in the pure testing phase (see details on topology metrics in Section~\ref{ssec:topological_assessment}).

Arrows in Table~\ref{table:metrics_summary} indicate whether each metric is better maximized or minimized.
Taking values between 0 and 1, DSC is a similarity metric that is better maximized ($\uparrow$). 
Difference metrics (ASSD, BNE1 and HR) must be minimized ($\downarrow$).

\begin{table*}[h!]
\small
  \centering
\begin{tabular}{  c   c c c } 
\toprule
  & Overlap & Boundary-distance & Topology \\ 
\midrule 
Training & DSC $\uparrow$ & ASSD $\downarrow$ & BNE1 $\downarrow$ \\ 
Testing & DSC $\uparrow$ & ASSD $\downarrow$ & HR $\downarrow$ \\ 
\bottomrule
\end{tabular}
\caption{Summary of the metrics used during the training phase (for learning monitoring) and testing phase (for evaluation). Arrows indicate whether higher $\uparrow$ or lower $\downarrow$ scores are better.}
\label{table:metrics_summary}
\end{table*}

\subsection{Experiments}
\label{ssec:experiments}

\subsubsection{$\mathcal{L}_{topo}$ parameter settings} 
\label{sssec:lambda_topo_settings}

Our first experiment consists in the setting of the \textit{TopoCP} $\lambda_{topo}$ parameter that quantifies the contribution of our topological loss.
As mentioned in the prior Section~\ref{sssec:training_strategy}, we use a cross-validation approach by means of which we assess multiple $\lambda_{topo}$ in order to determine an optimal value. The ideal $\lambda_{topo}$ is a dataset dependant hyperparameter.
According to~\cite{hu_topology-preserving_2019}, $\lambda_{topo}$ must be chosen to avoid the risk of over-weighting of $\mathcal{L}_{topo}$ over $\mathcal{L}_{bce}$. 
Indeed, while $\mathcal{L}_{bce}$ is defined at every voxel of the image, $\mathcal{L}_{topo}$ is only defined at some critical points. 
The values $0.0002$, $0.005$, $0.001$, $0.05$, $0.1$ and $0.2$ are the $N$ $\lambda_{topo}$ evaluated in the training phase. 
$TopoCP_{n}$ define the set of 4 networks trained for cross-validation with $\lambda_{n}$. We consider DSC, ASSD and BNE1 for evaluation. The average performances over the folds are computed for each $TopoCP_{n}$ networks and ranked for each metric. $TopoCP_{n}$ are finally ranked, based on the sum of metric-wise ranking, to elect the optimal $\lambda_{topo}$. The latter is then selected in the following experiments.

\subsubsection{Methods comparison}
\label{sssec:method_comparison}

We compare our \textit{TopoCP} method to the two reference segmentation methods \textit{Baseline} and \textit{Hybrid} on both the clinical and atlas test sets. We assess the three complementary metrics DSC, ASSD and HR.
We perform paired Wilcoxon rank-sum tests to assess the statistical significance between \textit{TopoCP} and the two reference configurations. Significance level is set to 0.05.

\subsubsection{GA analysis}
\label{sssec:ga_analysis}

The STA set presents a large and steady range of GA with one subject per week from 18 to 38 weeks of GA. We observe the quantitative performances of DSC, ASSD and HR along gestation, i.e. as a function of the subject GA.

\subsubsection{Spatial topological analysis}
\label{sssec:spatial_analysis}

\begin{figure}[h!]
    \centering
    \includegraphics[width=\linewidth]{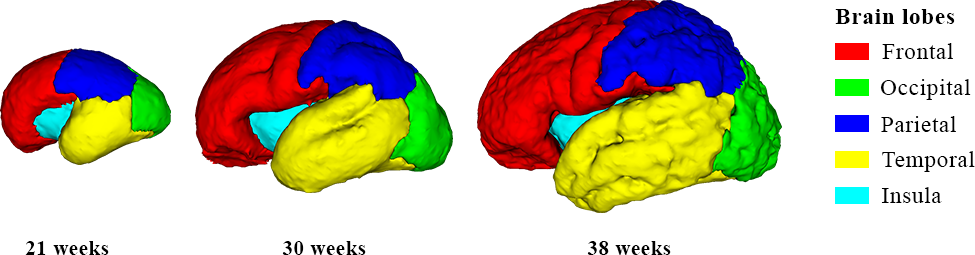} 
    \caption{3D rendering of the CP for STA subjects of 21, 30 and 38 weeks of gestation. The cortical volumes are split into the 5 lobes of the brain: the frontal lobe (red), the occipital lobe (green), the parietal lobe (dark blue), the temporal lobe (yellow), and the insula lobe (light blue). }
    \label{fig:Fig_lobes}
\end{figure}

From the STA set, we group the regional labels into 5 classes corresponding to the brain lobes, namely the frontal lobe, the occipital lobe, the parietal lobe, the temporal lobe, and the insula lobe. 
Figure~\ref{fig:Fig_lobes} shows 3D rendering of the finale maps of the brain lobes for the subject atlas of 21, 30 and 38 weeks of GA.

We proceed to a lobe-based analysis of the topology HR metric to analyze if \textit{TopoCP} benefits in one, some, all or none of them. We perform a paired Wilcoxon rank-sum test to assess the statistical significance between \textit{TopoCP} method and the reference configurations. Statistical ignificance level is set to 0.05.

\subsubsection{Group analysis: Neurotypical vs. Pathological}
\label{sssec:group_analysis}

The FeTA testing set presents a good heterogeneity in the neurotypical (N=4) and pathological (N=5) subjects. 
In each group, we observe the variation of the DSC, the ASSD and the HR with \textit{TopoCP}.
No statistical analysis is performed due to the small sample sizes (groups of N=4 and N=5).

\subsubsection{Manual annotations comparison}
\label{sssec:weak_labels_analysis}

Ultimately, we evaluate the performances of \textit{TopoCP} compared to the original FeTA manual annotations, using the topologically corrected segmentations as ground truth. 

Let us note that these original FeTA annotations are sparse and interpolated, hence resulting in \textit{noisy} references. Nevertheless, they are still used for training, as manual topological correction of 24 volumes would not be realistic (time/expertise effort). 
We quantitatively assess these segmentation with the DSC, the ASSD and the HR. The segmentation correctness difference was tested with the paired Wilcoxon rank-sum. The $p$-value level for statistical significance was set at 0.05.

\subsubsection{Experts evaluation}
\label{sssec:experts_evaluation}

Three experts in fetal brain MRI (two radiologists and one engineer) perform an independent and blind assessment of the three automatic segmentation methods on the clinical CHUV dataset. 
Each fetal brain MR exam is provided with the SR reconstructed volume, the subject's GA at scan time, the subject's group (i.e. Neurotypical or Pathological) and the three segmentation (from configurations \textit{Baseline}, \textit{Hybrid} and \textit{TopoCP}) that are randomly anonymized with labels $A$, $B$ and $C$.

The experts are asked to rank the segmentation masks $A$, $B$ and $C$ as Best, Medium and Worst.
Visualization of the images and their segmentation are done with the open-source ITK-SNAP~\cite{yushkevich_user-guided_2006} software. Specifically, binary segmentation are both visualized in 2D, as an overlay to the T2w gray-scale SR images, and in 3D, with the ITK-SNAP integrated 3D viewer.

We assess the inter-rater reliability with the percentage agreement and an ordinal Gwet's agreement coefficient (Gwet's AC) that we interpret according to Altman's benchmarking scale~\cite{gwet_handbook_2014}. We further consider a consensus evaluation as the majority voting of the experts' evaluation.

\section{Results and Discussion}
\label{sec:results_and_discussion}

\subsection{$\lambda_{topo}$ hyper-parameter tuning}
\label{ssec:results_cross_validation}

Table~\ref{tab:metrics_cv_topo} shows the averaged validation scores of all three configurations, and specifically for each $\lambda_{topo}$ assessed in the \textit{TopoCP} configuration.
Our first observation is that, regardless of the value of the $\lambda_{topo}$ parameter, \textit{TopoCP} is better performing than both reference methods, as we reach the state of the art performances in all the three complementary metrics.
Overall, all $TopoCP_{n}$ give similar DSC (mean: 0.76) and ASSD (mean $\pm$  standard deviation: 0.27 $\pm$ 0.01) performances, although $\lambda_{topo}=0.01$ is of the highest rank for both overlap and boundary-distance based scores.
An increased inter-$TopoCP_{n}$ variability appears for the topology-based metric (BNE$1$) with mean score from 20.6 to 22.5. 
Counting the number of bores in the CP segmentation, $\lambda_{topo}=0.005$ is giving the best performances. We observe large BNE$1$ standard deviation for all $TopoCP_{n}$. 
Nonetheless, the finest topology-relative $\lambda_{topo}$ is not only giving the minimum averaged BNE$1$, but is also noticeably presenting a smaller BNE$1$ standard deviation of 7.8 (BNE$1$ range: from 7.8 to 10). Therefore, $\lambda_{topo}=0.005$ is the most accurate and precise of the $\lambda_{topo}$ assessed as for the topology fidelity.
The substantial fluctuation in the topological metric shows the importance of the choice of the $\lambda_{topo}$ hyper-parameter.

Finally, our global ranking that is derived from metric-wise rankings evidences the ideal value $\lambda_{topo}=0.005$. 
We observe that none of the extreme values considered (i.e. $0.0002$ and $0.2$) are in the Top 3 best performing $\lambda_{topo}$. Therefore, we can say that although $\lambda_{topo}=0.005$ might not be the exact optimal $\lambda_{topo}$, it certainly falls in a relevant range and in the right order of magnitude.

\begin{table*}[h!]
    \centering
    \resizebox{\textwidth}{!}{
\begin{tabular}{cc ccc c }
\toprule
\multicolumn{2}{c}{Configuration}& DSC $\uparrow$ & ASSD $\downarrow$ & BNE$1$ $\downarrow$ & Ranking $\downarrow$ \\
\midrule

\multicolumn{2}{c}{\textbf{Baseline}}  & 0.748 $\pm$ 0.009 & 0.292 $\pm$ 0.02  &  29.8 $\pm$ 14.5 & \\
\multicolumn{2}{c}{\textbf{Hybrid}}    & 0.744 $\pm$ 0.004 & 0.297 $\pm$ 0.01  &  31 $\pm$ 13.4   &  \\

\midrule
\parbox[t]{3mm}{\multirow{5}{*}{\rotatebox[origin=c]{90}{\parbox[c]{2cm}{\centering \textbf{TopoCP} $\lambda_{topo}$ }}}}&
  0.0002  & 0.758 $\pm$ 0.007 (5) & 0.274 $\pm$ 0.01 (5)  &  22.1 $\pm$ 10 (4) & 5 \\
& 0.001   & 0.761 $\pm$ 0.007 (2) & 0.270 $\pm$ 0.01 (3) &  21.0 $\pm$ 8.5 (2) & 2 \\
& 0.005   & 0.760 $\pm$ 0.007 (3) & 0.269 $\pm$ 0.01 (2) &  \textbf{20.6 $\pm$ 7.8 (1)} & \textbf{1} \\
& 0.01    & \textbf{0.762 $\pm$ 0.007 (1)} & \textbf{0.268 $\pm$ 0.01 (1)}  &  22.5 $\pm$ 9.0 (5) & 2 \\
& 0.2     & 0.760 $\pm$ 0.005 (4) & 0.272 $\pm$ 0.01 (4) &  21.6 $\pm$ 7.9 (3) & 4 \\
\bottomrule
\end{tabular}}
\caption{Table of the validation scores (mean $\pm$ standard deviation) of the dice similarity coefficient (DSC), the average symmetric surface distance (ASSD) and the 1-dimensional Betti number error (BNE$1$). 
Arrows indicate whether the metric is better maximized $\uparrow$ or minimized $\downarrow$.
The best scores between all $\lambda_{topo}$ are shown in bold. 
A ranking for each metric is shown in parenthesis. The final ranking is formulated from the sum of metric-wise ranking scores. \textit{Baseline} stands for the $\mathcal{L}_{bce}$ loss and \textit{Hybrid} corresponds to $\mathcal{L}_{bce} + \mathcal{L}_{dice}$.}
\label{tab:metrics_cv_topo}
\end{table*}

\subsection{Methods comparison}
\label{ssec:results_methods_comparison}

\begin{figure}[h!]
    \centering
    \includegraphics[width=\linewidth]{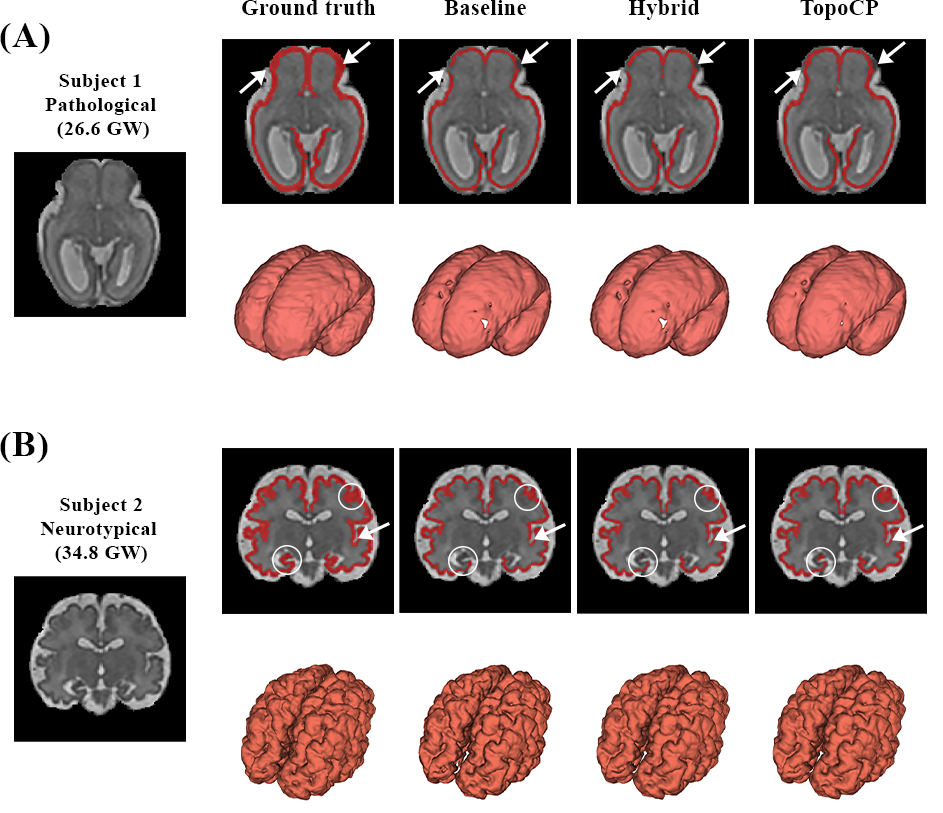} 
    \caption{ 
        Segmentation results on an axial view (top) and 3D rendering (bottom) of the cortical plate on FeTA subjects: (A) a neurotypical subject of 34.8 weeks of GA, and (B) a pathological subject of 28.1 weeks of GA. 
        Comparison of (a) the manually corrected ground truth segmentation, (b) the results of the \textit{Baseline} networks trained with $\mathcal{L}_{bce}$, (c) the results of the \textit{Hybrid} configuratio trained with $\mathcal{L}=\mathcal{L}_{bce}+\mathcal{L}_{dice}$ and (d) and the results obtained with our custom method \textit{TopoCP} trained with $\mathcal{L}=(1-\lambda_{topo})\mathcal{L}_{bce}+\lambda_{topo}\mathcal{L}_{topo}$, where $\lambda_{topo}=0.005$.
        White circles show representative area where cortical gyrification have a better in-depth segmentation with \textit{TopoCP} method.
        White arrows show fixed connections using our \textit{TopoCP} method compared to the reference segmentations.}
    \label{fig:Fig_feta_qualitative}
\end{figure}

Figure~\ref{fig:Fig_feta_qualitative} illustrates the accuracy of the fetal CP segmentation for a pathological subject of 26.6 (Subject 1) and a neurotypical subject of 34.8 (Subject 2) weeks of GA. The topologically corrected ground truth and the three configurations segmentation with \textit{Baseline}, \textit{Hybrid}, and \textit{TopoCP}, are compared. Qualitative 2D assessment (top rows) of the segmentation are presented as an overlay on the T2w image on an axial, respectively coronal, view for Subject 1, respectively Subject 2. Additionally, 3D rendering of the CP segmentation are presented in the bottom rows.
Overall, we observe that all configurations generates a thinner ribbon than the corrected ground truth. 
Specifically, \textit{TopoCP} presents fixed cortical connections that are broken in the \textit{Baseline} and \textit{Hybrid} segmentations (white arrows).
The CP \textit{TopoCP} segmentation 3D rendering seems to present less bores than the two reference configurations \textit{Baseline} and \textit{Hybrid}.
In particular, the segmentation appears, equivalently for the young and the old fetuses, more challenging, for all methods, in the lower parts of the frontal and temporal lobes, although \textit{TopoCP} seems to exhibit a more sensitive segmentation in these areas.
\textit{TopoCP} appears to be more sensitive to the complexity of the CP morphology in older fetuses. White circles evidence in Subject 2 an improved segmentation in the hippocampal area and the depth of a gyrification. 
White arrows show area where the topological correctness recovered with \textit{TopoCP}, compared to the \textit{Baseline}. 
In Subject 1, two connections are fixed in the frontal lobe, although one of them is already fixed in the \textit{Hybrid} configuration.

Quantitative results for both test sets (total of 27 cases), FeTA and STA, are respectively presented in Table~\ref{tab:test_quantitative_results_feta} and Table~\ref{tab:test_quantitative_results_sta}.
Tables show the mean ± standard deviation of the CP segmentation for each testing metric (DSC, ASSD and HR) in each configuration. 
Overall, the performance of the segmentation framework is improved when trained with our optimal \textit{TopoCP} configuration (i.e. $\lambda_{topo}=0.005$) for all metrics in both datasets. \textit{TopoCP} is always performing significantly better than both \textit{Baseline} and \textit{Hybrid} methods.

Furthermore, while all analyzed aspects of the segmentation, namely the overlap, the boundary-distance and the topology are improved with \textit{TopoCP} compared to the reference methods, we observe a drop in the performances between FeTA and STA evaluation. Indeed, DSC goes from 0.85 in FeTA to 0.79 in STA, ASSD from 0.19 to 0.41 and HR from 0.06 to 0.23. 
We believe this is due to the domain shift between FeTA and STA images (different reconstruction pipelines, different intensity-based processing, etc) generating an inter-dataset variation in the data distribution. Such domain gap between the two sets of images is not learned from the training data that are only composed of FeTA images. Still, DL can generalize to some extent. Further training with multi-dataset images or the use of domain adaptation strategies can partially fit the domain gap. Let us note that we do not address the data distribution generalization in this paper as it is beyond its scope. Furthermore, such performance drop occurs also in \textit{Baseline} and \textit{Hybrid} approach.

\begin{table}[h!]
    \begin{subtable}[h]{0.5\textwidth}
          \centering
\small
\resizebox{\linewidth}{!}{
\begin{tabular}{c  c   c   c}
\toprule
 & DSC $\uparrow$ & ASSD $\downarrow$ & HR $\downarrow$ \\
\midrule
\textit{Baseline}~    & \thead{0.82 $\pm$ 0.02} & \thead{ 0.22 $\pm$ 0.05} & \thead{0.093 $\pm$ 0.03 } \\
\textit{Hybrid}~      & \thead{0.82 $\pm$ 0.02} & \thead{ 0.23 $\pm$ 0.06} & \thead{0.10  $\pm$ 0.04 } \\
\textit{TopoCP}~      & \thead{\textbf{0.85 $\pm$ 0.01}\\\textbf{(*, $^{+}$)}} & \thead{ \textbf{0.19 $\pm$ 0.04}\\\textbf{(*, $^{+}$)}} & \thead{\textbf{0.06 $\pm$ 0.03}\\\textbf{(*, $^{+}$)}} \\
\bottomrule
\end{tabular}
}
       \caption{FeTA}
       \label{tab:test_quantitative_results_feta}
    \end{subtable}
    \hfill
    \begin{subtable}[h]{0.5\textwidth}
                  \centering
\small
\resizebox{\linewidth}{!}{
\begin{tabular}{c  c   c   c}
\toprule
 & DSC $\uparrow$ & ASSD $\downarrow$ & HR $\downarrow$ \\
\midrule
\textit{Baseline}~    & \thead{0.77 $\pm$ 0.05} & \thead{ 0.42 $\pm$ 0.14} & \thead{0.25 $\pm$ 0.10 } \\
\textit{Hybrid}~      & \thead{0.77 $\pm$ 0.05} & \thead{ 0.42 $\pm$ 0.15} & \thead{0.26 $\pm$ 0.11} \\
\textit{TopoCP}~      & \thead{\textbf{0.79 $\pm$ 0.05}\\\textbf{(*, $^{+}$)}} & \thead{\textbf{0.41 $\pm$ 0.18}\\\textbf{(*, $^{+}$)}} & \thead{\textbf{0.23 $\pm$ 0.10}\\\textbf{(*, $^{+}$)}} \\ 
\bottomrule
\end{tabular}
}
        \caption{STA}
        \label{tab:test_quantitative_results_sta}
     \end{subtable}
     \caption{Tables of the metrics computed on the pure testing sets FeTA~\ref{tab:test_quantitative_results_feta} and STA~\ref{tab:test_quantitative_results_sta}. Mean $\pm$ standard deviation for the dice similarity coefficient (DSC), Average symmetric surface distance (ASSD) and holes ratio (HR) are presented. 
     Arrows indicate whether the metric is better maximized $\uparrow$ or minimized $\downarrow$.
     The best scores between all three configurations are shown in bold. $p$-values of Wilcoxon rank sum test between \textit{TopoCP} and the reference configurations, \textit{Baseline} (*) and \textit{Hybrid} ($^{+}$), are considered statistically significant for $p<0.05$.
     }
     \label{tab:test_quantitative_results}
\end{table}

\subsection{Segmentation performance over gestation}
\label{ssec:results_f_GA}

\begin{figure}[h!]
    \centering
    \includegraphics[width=\linewidth]{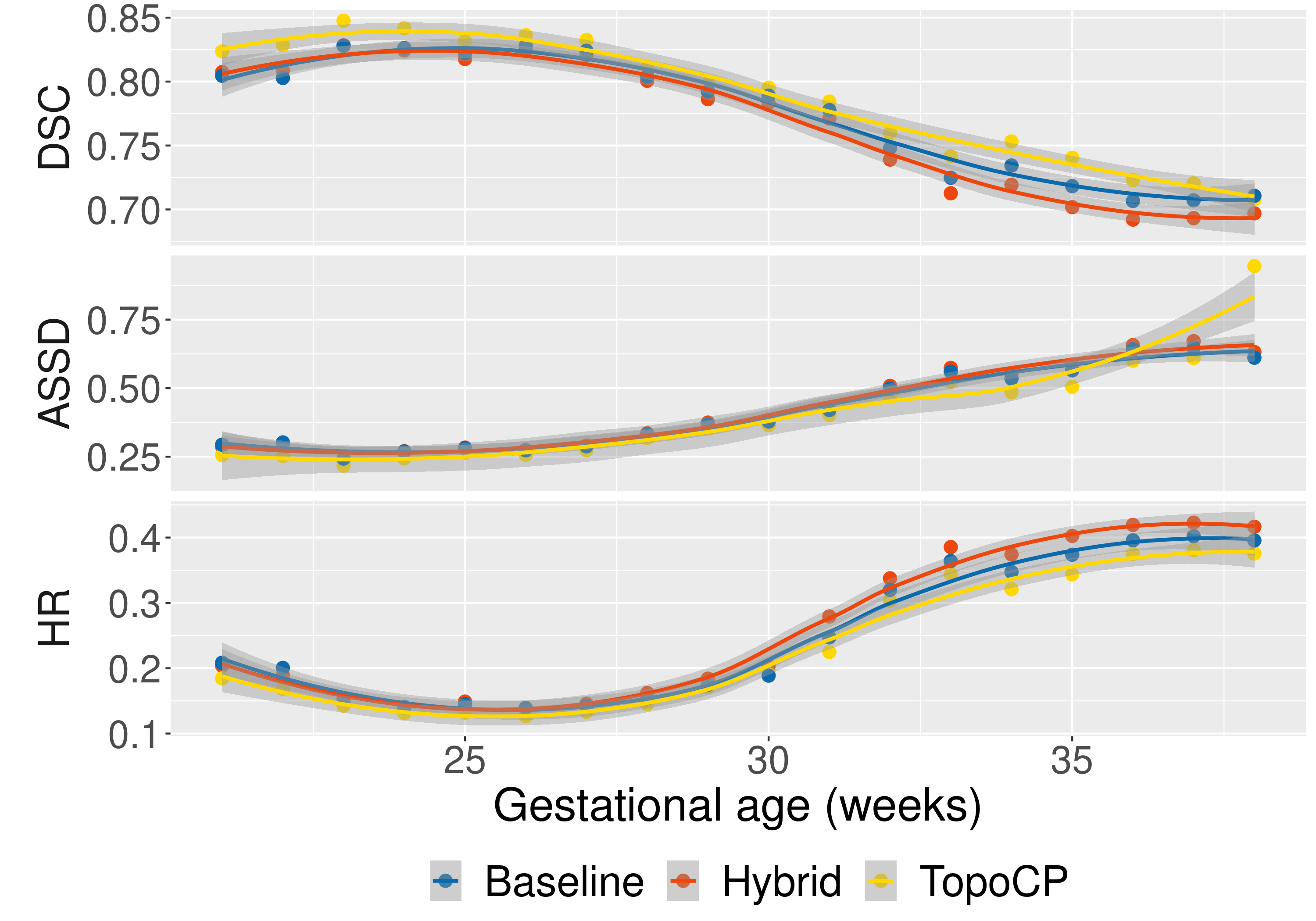} 
    \caption{Evolution of the performance metrics (DSC: Dice similarity coefficient; ASSD: average symmetric surface distance; HR: holes ratio) on the STA images as a function of the gestational age (from 21 to 38 weeks of gestation). }
    \label{fig:Fig_sta_metrics_f_GA}
\end{figure}

Taking advantage of the steady GA-distribution in the STA set, we perform an analysis of the metrics throughout gestation. Figure~\ref{fig:Fig_sta_metrics_f_GA} shows the performance metrics as a function of the GA for 18 cases from 21 to 37 weeks of GA.
Regardless of the configuration, we observe a trend in the performances based on the GA. Indeed, all metrics reach better performances for subjects younger than 30 weeks of GA. From week 23 to 31, all three methods (\textit{Basline}, \textit{Hybrid} and \textit{TopoCP}) seem to give equivalent scores. Outside this range (i.e. GA$<23$ and GA$>30$ weeks), \textit{TopoCP} is always performing better than the other two methods, except for one outlier subject of 38 weeks of GA.

We visually inspect the \textit{TopoCP} segmentation mask of the STA 38-weeks-old subject to better understand the origin of this outlier. Specifically in the cerebellum, we observe the presence of false positives that are connected to the main cortical segmentation through brainstem false positives. 
The cerebellum is a "little brain" composed of white matter encased in the cerebellar cortex. In terms of fetal brain T2w MR contrast and similarly to the CP, the cerebellar cortex expresses as a thin dark ribbon surrounding white matter. Therefore, it is a challenging area to accurately differentiate in the segmentation of the fetal CP at a patch-level.
Similar mis-segmentations appear in younger fetuses, nevertheless our post-processing step to keep the biggest connected component filters out most of it. In this oldest subject, cerebellum errors are worsened with brainstem mis-segmentation. We hypothesize such false positive errors in the cerebellum are due to mis-leading contextual information, due to the reduced field of view of the patches. Therefore, we believe that increasing the patch size could help to overcome these mis-segmentation. 
Nevertheless, while the performances are particularly damaged in the distance metric, \textit{TopoCP} still performs better in terms of DSC and HR compares to the other configurations. 

Overall, it is in the second half of the third trimester GAs (i.e. GA$>$30 weeks) that we observe an increased benefit from \textit{TopoCP}, compared to other methods. Our topological loss has a stronger positive effect on the topological errors for old subjects with more complex topology, although the whole range of gestational age consistently presents benefits from \textit{TopoCP}.

We derive two hypothesis on the variation of the performances throughout gestation. First, we recall that the training data present subjects in the range 20.9 to 34.8 weeks of GA with mean 28.2 and standard deviation 3.6. Therefore, variation of young and old fetal brains are less represented in the learning process. Additionally, third trimester subjects present advanced sulcal patterns, resulting in a substantially more complex topology.
Therefore, we postulate this accentuate the unstable evolution of segmentation accuracy over gestation.

\subsection{Topology analysis per brain lobes}
\label{ssec:results_f_lobe}

\begin{figure}[h!]
    \centering
    \includegraphics[width=\linewidth]{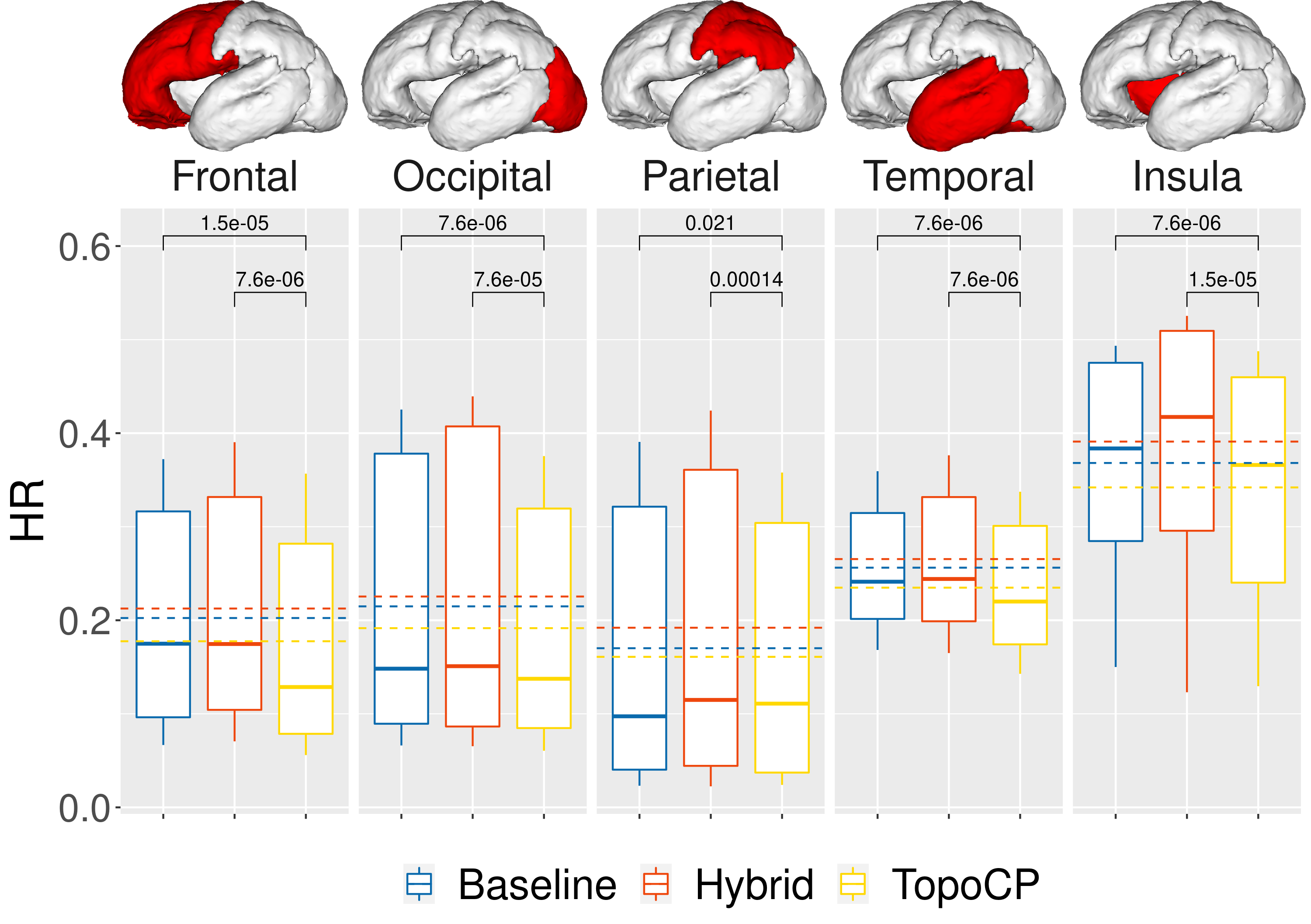} 
    \caption{Comparison of the hole ratio (HR) in the fetal brain lobes (frontal, occipital, parietal, temporal and insula) on the STA dataset for all configurations (\textit{Baseline}, \textit{Hybrid} and \textit{TopoCP}).
    Dashed horizontal lines indicate the per-lobe mean HR for each configuration.
    \textit{p}-values of paired Wilcoxon rank-sum tests are displayed comparing \textit{TopoCP} to each reference methods. }
    \label{fig:Fig_sta_per_region}
\end{figure}

Figure~\ref{fig:Fig_sta_per_region} presents a comparison of HR at a lobe-level between the configurations.
This boxplot evidences the significant benefits ($p<0.05$) of \textit{TopoCP} in most areas (frontal, occipital, temporal and insula lobes) compared to the \textit{Baseline} configuration. In the parietal lobe, \textit{TopoCP} is on average performing better than the \textit{Baseline} although without statistical significance. Compared to the \textit{Hybrid} configuration, \textit{TopoCP} presents a significantly lower HR in all brain lobes. 
Regardless of the configuration, the parietal lobe is always the better segmented lobe in terms of HR as opposed to the insula lobe.

\subsection{Groups analysis}
\label{ssec:results_f_group}

Table~\ref{tab:metrics_per_group} summarizes the performance metrics (mean $\pm$ standard deviation) in both neurotypical (NT) and pathological (PT) groups. We observe the same trend in this group-based analysis as in the overall method comparison (see Tables~\ref{tab:test_quantitative_results} in Section~\ref{ssec:results_methods_comparison}).
Regrettably, precise information on the pathology are not available in the FeTA dataset meta data. Therefore, we cannot draw any conclusion relative to eventual cortical pathologies.

\begin{table*}[h!]
  \centering
\resizebox{\linewidth}{!}{
\begin{tabular}{c   c   c   c   c   c   c}
\toprule
 & \multicolumn{2}{c}{DSC $\uparrow$} & \multicolumn{2}{c}{ASSD $\downarrow$} & \multicolumn{2}{c}{HR $\downarrow$} \\
      & NT & PT & NT & PT & NT & PT \\ 
\midrule
\textit{Baseline}~    & ~0.82 $\pm$ 0.01 & 0.82 $\pm$ 0.02 & 0.24 $\pm$ 0.07 & 0.21 $\pm$ 0.03 & 0.09 $\pm$ 0.04 & 0.09 $\pm$ 0.04  \\
\textit{Hybrid}~      & ~0.81 $\pm$ 0.02 & 0.82 $\pm$ 0.02 & 0.25 $\pm$ 0.07 & 0.22 $\pm$ 0.04 & 0.1  $\pm$ 0.05 & 0.1  $\pm$ 0.04  \\
\textit{TopoCP}~      & ~\textbf{0.84 $\pm$ 0.01}~ & ~\textbf{0.85 $\pm$ 0.01}~ & ~\textbf{0.21 $\pm$ 0.06}~ & ~\textbf{0.18 $\pm$ 0.02}~ & ~\textbf{0.06 $\pm$ 0.03}~ & ~\textbf{0.06 $\pm$ 0.03}~  \\
\bottomrule
\end{tabular} 
}
\caption{\label{tab:metrics_per_group}
Table of the metrics computed on the FeTA pure testing for neurotypical (NT) and pathological (PT) groups. Mean $\pm$ standard deviation for the dice similarity coefficient (DSC), Average symmetric surface distance (ASSD) and hole ratio (HR) are presented. 
Arrows indicate whether the metric is better maximized $\uparrow$ or minimized $\downarrow$.
The best scores between our \textit{TopoCP} method and the original annotations are shown in bold.}
\end{table*}

\subsection{Robustness to noisy manual annotations}
\label{ssec:results_weak_labels}

\begin{figure}[h!]
    \centering
    \includegraphics[width=\linewidth]{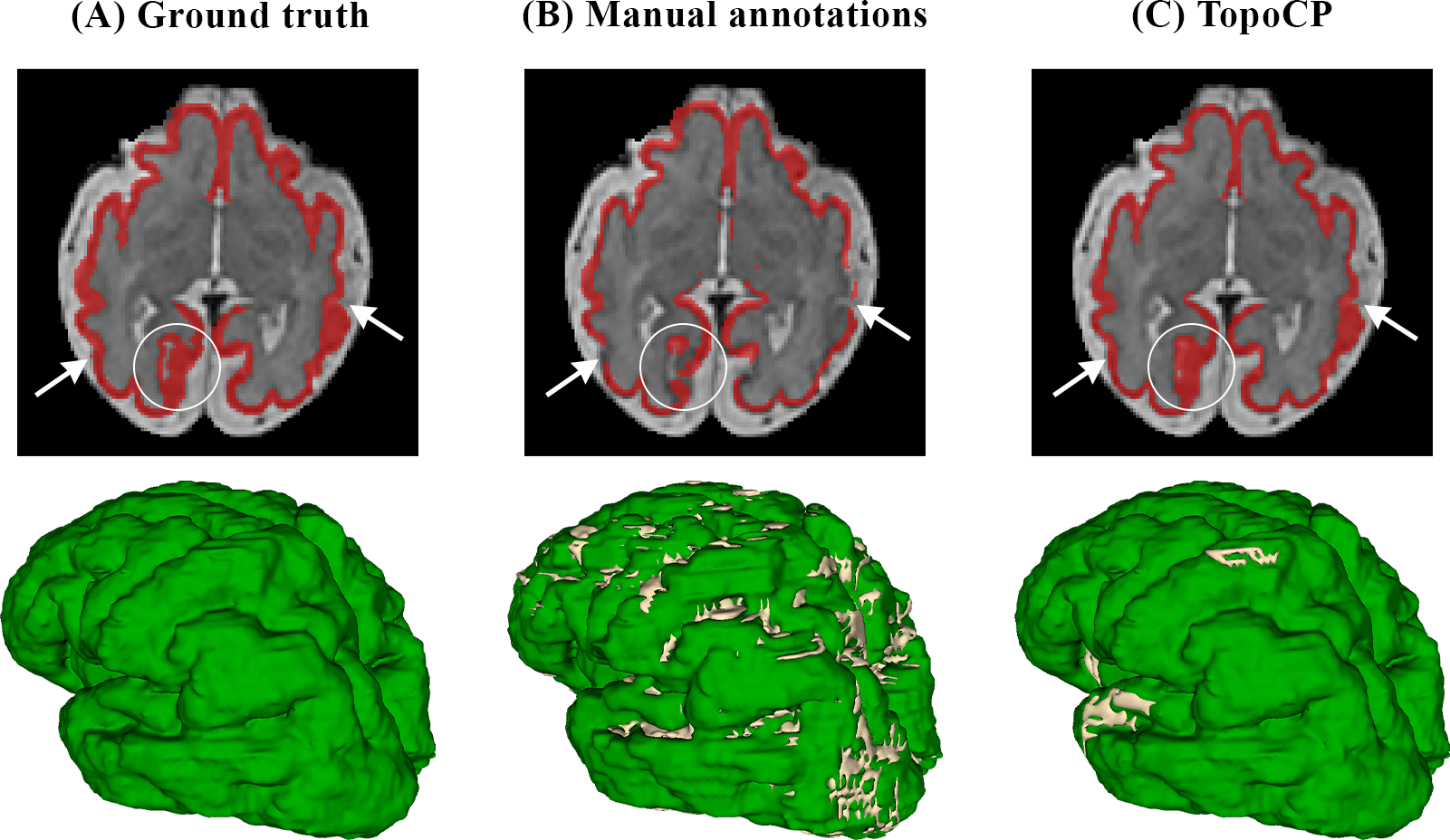} 
    \caption{Qualitative 2D and 3D assessment of CP segmentation on a 31.5 weeks of GA neurotypical subject.
    Comparison of (A) the corrected ground truth to (B) the original manual annotation and (C) our \textit{TopoCP} automatic segmentation method. Segmentations are overlaid on a T2w axial view (top). 
    Segmentation 3D renderings (bottom) highlight the true positives (green) and false negatives connected to 1-dimensional holes (light red).}
    \label{fig:Fig_manual_vs_topo}
\end{figure}

Figure~\ref{fig:Fig_manual_vs_topo} (top) shows a comparative T2w axial view of the ground truth topologically corrected segmentation (A), the original manual annotation provided in FeTA (B) and the \textit{TopoCP} predicted segmentation (C). Overall, we observe an improved accuracy with the automatic method. Specifically, white arrows indicate cortical location where \textit{TopoCP} fixes topological inconsistencies, compared to the original manual annotations. The white circle focuses on the hippocampal area where the manual annotations are confused. 
Figure~\ref{fig:Fig_manual_vs_topo} (bottom) shows 3D rendering of the true cortical volume (green). In the manual and automatic \textit{TopoCP} segmentation, the 1-dimensional holes are evidenced with the false negatives connected to 1-dimensional holes (light camel).

Table~\ref{tab:metrics_manual_vs_topo} shows a comparison of the performance metrics of the original FeTA manual annotations on 9 subjects and our \textit{TopoCP} method. 
\textit{TopoCP} is significantly better (*) than the original manual annotations in all metrics (DSC, ASSD and HR). 

The automatic \textit{TopoCP} segmentation method is able to learn segmentation features from noisy annotations. This improvement is conveyed in all three similarity, boundary-distance and topology -based metrics.

\begin{table*}[h!]
  \centering
\begin{tabular}{c   c c  c c  c c}
\toprule
 & \multicolumn{2}{c}{DSC $\uparrow$} & \multicolumn{2}{c}{ASSD $\downarrow$} & \multicolumn{2}{c}{HR $\downarrow$} \\
\midrule
\textit{TopoCP}~~      & \textbf{0.85 $\pm$ 0.01} & \multirow{2}{*}{(*)~} & \textbf{0.19 $\pm$ 0.04} &  \multirow{2}{*}{(*)~} & \textbf{0.062 $\pm$ 0.03} &  \multirow{2}{*}{(*)~}   \\
\textit{Manual}~~      & 0.82 $\pm$ 0.02 & & 0.23 $\pm$ 0.04 & & 0.23 $\pm$ 0.06 &   \\
\bottomrule
\end{tabular}
\caption{\label{tab:metrics_manual_vs_topo}
Table of the metrics computed on the pure testing sets FeTA. Mean $\pm$ standard deviation for the dice similarity coefficient (DSC), Average symmetric surface distance (ASSD) and hole ratio (HR) are presented. 
Arrows indicate whether the metric is better maximized $\uparrow$ or minimized $\downarrow$.
The best scores between our \textit{TopoCP} method and the original annotations are shown in bold. $p$-values of Wilcoxon rank sum test between \textit{TopoCP} and the original annotations are considered statistically significant (*) for $p<0.05$.}
\end{table*}

\subsection{Out-of-domain qualitative assessment}
\label{ssec:results_qualitative_chuv}

Table~\ref{tab:experts_qualitative_results} summarizes the classification results of the segmentation masks according to each expert into Worst, Medium or Best. A consensus of the three experts assessment is presented in the bottom row.

The estimated Gwet's AC between the three experts was $0.68$ (standard error (SE): $0.10$) for Worst,  $0.68$ (SE: $0.10$) for Medium and $1$ for Best segmentation classifications.
According to Altman's benchmarking scale, the estimated coefficients for Worst and Medium segmentations are considered to be either \textit{Moderate}, \textit{Good} or \textit{Very Good} with a probability of $0.99$. The Best segmentation classification has a \textit{Very good} agreement between the experts with a probability of 1. 
With a percentage agreement of 100 \%, the consensus of experts classifies \textit{TopoCP} as the Best segmentation method in 100\% of the cases. Inter-rater discrepancies are observed in the choice of Worst and Medium between \textit{Baseline} and \textit{Hybrid} segmentation.

Although \textit{TopoCP} is ranked as Best segmentation in all cases, predictions still present many segmentation errors. We emphasize that the distribution of this clinical set differs from the FeTA training set as they were generated with different SR methods. Nevertheless, while all three configurations present altered segmentation due to the domain shift, still, \textit{TopoCP} remains the better performing method.

\begin{table*}[h!]
    \centering
    \resizebox{\textwidth}{!}{
\begin{tabular}{c ccc ccc ccc }
\toprule
 & \multicolumn{3}{c}{Best} & \multicolumn{3}{c}{Medium} & \multicolumn{3}{c}{Worst} \\%
\cmidrule(lr){2-4}\cmidrule(lr){5-7}\cmidrule(lr){8-10}

 & \textit{Baseline} & \textit{Hybrid}  &  \textit{TopoCP}   & \textit{Baseline} & \textit{Hybrid}  &  \textit{TopoCP}   & \textit{Baseline} & \textit{Hybrid}  &  \textit{TopoCP} \\
\midrule
Radiologist 1   & 0 & 0 & 31 & 25 & 6 & 0 & 6 & 25 & 0  \\
Radiologist 2   & 0 & 0 & 31 & 26 & 5 & 0 & 5 & 26 & 0  \\
Engineer        & 0 & 0 & 31 & 23 & 8 & 0 & 8 & 23 & 0  \\
\midrule

Percentage agreement   & \multicolumn{3}{c}{100 \%} & \multicolumn{3}{c}{78 \%} & \multicolumn{3}{c}{78 \%} \\
\thead{Gwet's AC (SE)}   & \multicolumn{3}{c}{1 (-)} & \multicolumn{3}{c}{0.68 (0.10)} & \multicolumn{3}{c}{0.68 (0.10)} \\
\cmidrule(lr){1-1} \cmidrule(lr){2-10}
Consensus   & 0 & 0 & \textbf{31} & 26 & 5 & 0 & 5 & 26 & 0  \\
\bottomrule
\end{tabular}}
\caption{Table of the three experts (two radiologists and one engineer) qualitative ranking of the three segmentation configurations (\textit{Baseline}, \textit{Hybrid} and \textit{TopoCP}) as Best, Medium and Worst. Percentage agreement between experts and Gwet's AC with standard error (SE) are presented for each ranking category. Finally, a consensus ranking is presented in the bottom row as a majority voting of the experts' evaluation.}
\label{tab:experts_qualitative_results}
\end{table*}

\section{Conclusion}
\label{sec:conclusion}


In this work, we developed a topological loss function for the optimization of deep-learning based segmentation methods of the fetal cortical plate in MRI. Our core contribution lies in the multi-dimensional approach of this generalized loss function. Jointly, we presented an original topology-based metric to quantify the 1-dimensional topological errors both in terms of count and size. 
We presented extensive quantitative and qualitative validation on a total of 58 fetal brains of a wide range of GA (from 21 to 38 weeks of GA), including both neurotypical and pathological subjects.
We compared our \textit{TopoCP} method to (i) state-of-the-art methods and (ii) semi-automatic \textit{noisy} reference segmentation. 
Experiments have shown that the integration of a topological constraint in the segmentation framework of the CP in fetal brain MRI significantly benefits not only the shape correctness - as it first aims, but also the overlap and distance accuracy. 
Although our segmentation framework is implemented for 2D image patches, 3D information is integrated thanks to the multi-view pipeline with the extraction of patches from the three orthogonal orientations (axial, coronal and sagittal). While our approach cannot be considered to be 3D, yet the benefit of our multi-dimensional topological loss is conveyed in the 3D metrics, including the topology-based one. Nevertheless, we believe that the adoption of a real 3D-based framework could only improve the overall performances, although we acknowledge that the computational cost of the topological loss in this process is an important shortcoming. 
Moreover, results evidence that the generalization of the learned topology is not hampered by the \textit{noisiness} of the manual annotations used for training.

This study is the first to address both the specific improvement of the topological correctness of the CP segmentation, and the definition of a topological assessment. 
The reduced gap in the topological and shape correctness accuracy is ultimately associated with minimal manual refinement needed for further quantitative surface-based analysis. 
Future work will focus on a wider generalization of our method application. Indeed, while our method is formulated to consider multiple dimensions, we only present a 2D application. The overall framework can be generalized for a 3D-based approach. Similarly, we focused here on a single-tissue, namely the cortical GM segmentation, segmentation problem, although generalization to a multi-tissue segmentation approach could be applied.

\section{Acknowledgments}
\label{sec:acknowledgments}

This work is supported by the Swiss National Science Foundation through grants 182602 and 141283. We acknowledge access to the facilities and expertise of the CIBM Center for Biomedical Imaging, a Swiss research center of excellence founded and supported by Lausanne University Hospital (CHUV), University of Lausanne (UNIL), Ecole polytechnique fédérale de Lausanne (EPFL), University of Geneva (UNIGE) and Geneva University Hospitals (HUG).
We acknowledge Dr Hélène Lajous and Andrés Le Boeuf for their help in the topology correction of the manual annotations.

The authors have no relevant financial or non-financial interests to disclose.

%
%
\clearpage
\bibliography{references}

\begin{thebibliography}{}

\bibitem[noa, ]{noauthor_gudhi_nodate}
{GUDHI}, {Geometry} {Understanding} in {Higher} {Dimensions}.

\bibitem[Barkovich et~al., 2005]{barkovich_developmental_2005}
Barkovich, A.~J., Kuzniecky, R.~I., Jackson, G.~D., Guerrini, R., and Dobyns,
  W.~B. (2005).
\newblock A developmental and genetic classification for malformations of
  cortical development.
\newblock {\em Neurology}, 65(12):1873--1887.

\bibitem[Caldairou et~al., 2011]{caldairou_segmentation_2011}
Caldairou, B., Passat, N., Habas, P., Studholme, C., Koob, M., Dietemann,
  J.-L., and Rousseau, F. (2011).
\newblock Segmentation of the cortex in fetal {MRI} using a topological model.
\newblock In {\em 2011 {IEEE} {International} {Symposium} on {Biomedical}
  {Imaging}: {From} {Nano} to {Macro}}, pages 2045--2048, Chicago, IL, USA.
  IEEE.

\bibitem[Clouchoux et~al., 2013]{clouchoux_delayed_2013}
Clouchoux, C., du~Plessis, A.~J., Bouyssi-Kobar, M., Tworetzky, W., McElhinney,
  D.~B., Brown, D.~W., Gholipour, A., Kudelski, D., Warfield, S.~K., McCarter,
  R.~J., Robertson, R.~L., Evans, A.~C., Newburger, J.~W., and Limperopoulos,
  C. (2013).
\newblock Delayed {Cortical} {Development} in {Fetuses} with {Complex}
  {Congenital} {Heart} {Disease}.
\newblock {\em Cerebral Cortex}, 23(12):2932--2943.

\bibitem[Clouchoux et~al., 2012]{clouchoux_normative_2012}
Clouchoux, C., Guizard, N., Evans, A.~C., du~Plessis, A.~J., and Limperopoulos,
  C. (2012).
\newblock Normative fetal brain growth by quantitative in vivo magnetic
  resonance imaging.
\newblock {\em American Journal of Obstetrics and Gynecology},
  206(2):173.e1--173.e8.

\bibitem[Clough et~al., 2021]{clough_topological_2021}
Clough, J., Byrne, N., Oksuz, I., Zimmer, V.~A., Schnabel, J.~A., and King, A.
  (2021).
\newblock A {Topological} {Loss} {Function} for {Deep}-{Learning} based {Image}
  {Segmentation} using {Persistent} {Homology}.
\newblock {\em IEEE Transactions on Pattern Analysis and Machine Intelligence},
  pages 1--1.

\bibitem[Cohen-Steiner et~al., 2010]{cohen-steiner_lipschitz_2010}
Cohen-Steiner, D., Edelsbrunner, H., Harer, J., and Mileyko, Y. (2010).
\newblock Lipschitz {Functions} {Have} {L} p -{Stable} {Persistence}.
\newblock {\em Foundations of Computational Mathematics}, 10(2):127--139.

\bibitem[Dice, 1945]{dice_measures_1945}
Dice, L.~R. (1945).
\newblock Measures of the {Amount} of {Ecologic} {Association} {Between}
  {Species}.
\newblock {\em Ecology}, 26(3):297--302.

\bibitem[Dou et~al., 2021]{dou_deep_2021}
Dou, H., Karimi, D., Rollins, C.~K., Ortinau, C.~M., Vasung, L., Velasco-Annis,
  C., Ouaalam, A., Yang, X., Ni, D., and Gholipour, A. (2021).
\newblock A {Deep} {Attentive} {Convolutional} {Neural} {Network} for
  {Automatic} {Cortical} {Plate} {Segmentation} in {Fetal} {MRI}.
\newblock {\em IEEE Transactions on Medical Imaging}, 40(4):1123--1133.

\bibitem[Ebner et~al., 2020]{ebner_automated_2020}
Ebner, M., Wang, G., Li, W., Aertsen, M., Patel, P.~A., Aughwane, R.,
  Melbourne, A., Doel, T., Dymarkowski, S., De~Coppi, P., David, A.~L.,
  Deprest, J., Ourselin, S., and Vercauteren, T. (2020).
\newblock An automated framework for localization, segmentation and
  super-resolution reconstruction of fetal brain {MRI}.
\newblock {\em NeuroImage}, 206:116324.

\bibitem[Egaña-Ugrinovic et~al., 2013]{egana-ugrinovic_differences_2013}
Egaña-Ugrinovic, G., Sanz-Cortes, M., Figueras, F., Bargalló, N., and
  Gratacós, E. (2013).
\newblock Differences in cortical development assessed by fetal {MRI} in
  late-onset intrauterine growth restriction.
\newblock {\em American Journal of Obstetrics and Gynecology},
  209(2):126.e1--126.e8.

\bibitem[Fetit et~al., 2020]{fetit_deep_2020}
Fetit, A., Alansary, A., Cordero-Grande, L., Cupitt, J., Davidson, A., Edwards,
  A., Hajnal, J., Hughes, E., Kamnitsas, K., Kyriakopoulou, V., Makropoulos,
  A., Patkee, P., Price, A., Rutherford, M., and Rueckert, D. (2020).
\newblock A deep learning approach to segmentation of the developing cortex in
  fetal brain {MRI} with minimal manual labeling.
\newblock pages 241--261. PMLR.

\bibitem[Garcia et~al., 2018]{garcia_mechanics_2018}
Garcia, K.~E., Kroenke, C.~D., and Bayly, P.~V. (2018).
\newblock Mechanics of cortical folding: stress, growth and stability.
\newblock {\em Philosophical Transactions of the Royal Society B: Biological
  Sciences}, 373(1759):20170321.

\bibitem[Garel, 2004]{garel_mri_2004}
Garel, C. (2004).
\newblock {\em {MRI} of the {Fetal} {Brain}}.
\newblock Springer Berlin Heidelberg, Berlin, Heidelberg.

\bibitem[Garel et~al., 2003]{garel_fetal_2003}
Garel, C., Elmaleh, M., Chantrel, E., Sebag, G., and Brisse, H. (2003).
\newblock Fetal {MRI}: normal gestational landmarks for cerebral biometry,
  gyration and myelination.
\newblock {\em Child's Nervous System}, 19(7-8):422--425.

\bibitem[Gholipour et~al., 2014]{gholipour_fetal_2014}
Gholipour, A., Estroff, J.~A., Barnewolt, C.~E., Robertson, R.~L., Grant,
  P.~E., Gagoski, B., Warfield, S.~K., Afacan, O., Connolly, S.~A., Neil,
  J.~J., Wolfberg, A., and Mulkern, R.~V. (2014).
\newblock Fetal {MRI}: {A} {Technical} {Update} with {Educational}
  {Aspirations}.
\newblock {\em Concepts in Magnetic Resonance. Part A, Bridging Education and
  Research}, 43(6):237--266.

\bibitem[Gholipour et~al., 2010]{gholipour_robust_2010}
Gholipour, A., Estroff, J.~A., and Warfield, S.~K. (2010).
\newblock Robust {Super}-{Resolution} {Volume} {Reconstruction} {From} {Slice}
  {Acquisitions}: {Application} to {Fetal} {Brain} {MRI}.
\newblock {\em IEEE Transactions on Medical Imaging}, 29(10):1739--1758.

\bibitem[Gholipour et~al., 2017]{gholipour_normative_2017}
Gholipour, A., Rollins, C.~K., Velasco-Annis, C., Ouaalam, A., Akhondi-Asl, A.,
  Afacan, O., Ortinau, C.~M., Clancy, S., Limperopoulos, C., Yang, E., Estroff,
  J.~A., and Warfield, S.~K. (2017).
\newblock A normative spatiotemporal {MRI} atlas of the fetal brain for
  automatic segmentation and analysis of early brain growth.
\newblock {\em Scientific Reports}, 7(1):476.

\bibitem[Griffiths et~al., 2010]{griffiths_prospective_2010}
Griffiths, P., Reeves, M., Morris, J., Mason, G., Russell, S., Paley, M., and
  Whitby, E. (2010).
\newblock A {Prospective} {Study} of {Fetuses} with {Isolated}
  {Ventriculomegaly} {Investigated} by {Antenatal} {Sonography} and {In}
  {Utero} {MR} {Imaging}.
\newblock {\em American Journal of Neuroradiology}, 31(1):106--111.

\bibitem[Griffiths et~al., 2019]{griffiths_mri_2019}
Griffiths, P.~D., Bradburn, M., Campbell, M.~J., Cooper, C.~L., Embleton, N.,
  Graham, R., Hart, A.~R., Jarvis, D., Kilby, M.~D., Lie, M., Mason, G.,
  Mandefield, L., Mooney, C., Pennington, R., Robson, S.~C., and Wailoo, A.
  (2019).
\newblock {MRI} in the diagnosis of fetal developmental brain abnormalities:
  the {MERIDIAN} diagnostic accuracy study.
\newblock {\em Health Technology Assessment (Winchester, England)},
  23(49):1--144.

\bibitem[Gwet, 2014]{gwet_handbook_2014}
Gwet, K.~L. (2014).
\newblock {\em Handbook of inter-rater reliability: the definitive guide to
  measuring the extent of agreement among raters}.
\newblock Advances Analytics, LLC, Gaithersburg, Md, fourth edition edition.

\bibitem[Hong et~al., 2020]{hong_fetal_2020}
Hong, J., Yun, H.~J., Park, G., Kim, S., Laurentys, C.~T., Siqueira, L.~C.,
  Tarui, T., Rollins, C.~K., Ortinau, C.~M., Grant, P.~E., Lee, J.-M., and Im,
  K. (2020).
\newblock Fetal {Cortical} {Plate} {Segmentation} {Using} {Fully}
  {Convolutional} {Networks} {With} {Multiple} {Plane} {Aggregation}.
\newblock {\em Frontiers in Neuroscience}, 14:591683.

\bibitem[Hu et~al., 2019]{hu_topology-preserving_2019}
Hu, X., Li, F., Samaras, D., and Chen, C. (2019).
\newblock Topology-{Preserving} {Deep} {Image} {Segmentation}.
\newblock In Wallach, H., Larochelle, H., Beygelzimer, A., Alché-Buc, F.~d.,
  Fox, E., and Garnett, R., editors, {\em Advances in {Neural} {Information}
  {Processing} {Systems}}, volume~32. Curran Associates, Inc.

\bibitem[Im et~al., 2013]{im_quantification_2013}
Im, K., Pienaar, R., Paldino, M.~J., Gaab, N., Galaburda, A.~M., and Grant,
  P.~E. (2013).
\newblock Quantification and discrimination of abnormal sulcal patterns in
  polymicrogyria.
\newblock {\em Cerebral Cortex (New York, N.Y.: 1991)}, 23(12):3007--3015.

\bibitem[Khalili et~al., 2019]{khalili_automatic_2019}
Khalili, N., Lessmann, N., Turk, E., Claessens, N., Heus, R.~d., Kolk, T.,
  Viergever, M., Benders, M., and Išgum, I. (2019).
\newblock Automatic brain tissue segmentation in fetal {MRI} using
  convolutional neural networks.
\newblock {\em Magnetic Resonance Imaging}, 64:77--89.

\bibitem[Khawam et~al., 2021]{khawam_fetal_2021}
Khawam, M., de~Dumast, P., Deman, P., Kebiri, H., Yu, T., Tourbier, S., Lajous,
  H., Hagmann, P., Maeder, P., Thiran, J.-P., Meuli, R., Dunet, V.,
  Bach~Cuadra, M., and Koob, M. (2021).
\newblock Fetal {Brain} {Biometric} {Measurements} on {3D} {Super}-{Resolution}
  {Reconstructed} {T2}-{Weighted} {MRI}: {An} {Intra}- and {Inter}-observer
  {Agreement} {Study}.
\newblock {\em Frontiers in Pediatrics}, 9:639746.

\bibitem[Kuklisova-Murgasova et~al.,
  2012]{kuklisova-murgasova_reconstruction_2012}
Kuklisova-Murgasova, M., Quaghebeur, G., Rutherford, M.~A., Hajnal, J.~V., and
  Schnabel, J.~A. (2012).
\newblock Reconstruction of fetal brain {MRI} with intensity matching and
  complete outlier removal.
\newblock {\em Medical Image Analysis}, 16(8):1550--1564.

\bibitem[Kyriakopoulou et~al., 2017]{kyriakopoulou_normative_2017}
Kyriakopoulou, V., Vatansever, D., Davidson, A., Patkee, P., Elkommos, S.,
  Chew, A., Martinez-Biarge, M., Hagberg, B., Damodaram, M., Allsop, J., Fox,
  M., Hajnal, J.~V., and Rutherford, M.~A. (2017).
\newblock Normative biometry of the fetal brain using magnetic resonance
  imaging.
\newblock {\em Brain Structure and Function}, 222(5):2295--2307.

\bibitem[Leibovitz et~al., 2022]{leibovitz_fetal_2022}
Leibovitz, Z., Lerman-Sagie, T., and Haddad, L. (2022).
\newblock Fetal {Brain} {Development}: {Regulating} {Processes} and {Related}
  {Malformations}.
\newblock {\em Life}, 12(6):809.

\bibitem[Lenroot and Giedd, 2006]{lenroot_brain_2006}
Lenroot, R.~K. and Giedd, J.~N. (2006).
\newblock Brain development in children and adolescents: {Insights} from
  anatomical magnetic resonance imaging.
\newblock {\em Neuroscience \& Biobehavioral Reviews}, 30(6):718--729.

\bibitem[Leventer et~al., 2008]{leventer_malformations_2008}
Leventer, R.~J., Guerrini, R., and Dobyns, W.~B. (2008).
\newblock Malformations of cortical development and epilepsy.
\newblock {\em Dialogues in Clinical Neuroscience}, 10(1):47--62.

\bibitem[Maier-Hein et~al., 2022]{maier-hein_metrics_2022}
Maier-Hein, L., Reinke, A., Christodoulou, E., Glocker, B., Godau, P., Isensee,
  F., Kleesiek, J., Kozubek, M., Reyes, M., Riegler, M.~A., Wiesenfarth, M.,
  Baumgartner, M., Eisenmann, M., Heckmann-Nötzel, D., Kavur, A.~E., Rädsch,
  T., Tizabi, M.~D., Acion, L., Antonelli, M., Arbel, T., Bakas, S., Bankhead,
  P., Benis, A., Cardoso, M.~J., Cheplygina, V., Cimini, B., Collins, G.~S.,
  Farahani, K., van Ginneken, B., Hashimoto, D.~A., Hoffman, M.~M., Huisman,
  M., Jannin, P., Kahn, C.~E., Karargyris, A., Karthikesalingam, A., Kenngott,
  H., Kopp-Schneider, A., Kreshuk, A., Kurc, T., Landman, B.~A., Litjens, G.,
  Madani, A., Maier-Hein, K., Martel, A.~L., Mattson, P., Meijering, E., Menze,
  B., Moher, D., Moons, K. G.~M., Müller, H., Nickel, F., Nichyporuk, B.,
  Petersen, J., Rajpoot, N., Rieke, N., Saez-Rodriguez, J., Gutiérrez, C.~S.,
  Shetty, S., van Smeden, M., Sudre, C.~H., Summers, R.~M., Taha, A.~A.,
  Tsaftaris, S.~A., Van~Calster, B., Varoquaux, G., and Jäger, P.~F. (2022).
\newblock Metrics reloaded: {Pitfalls} and recommendations for image analysis
  validation.
\newblock Publisher: arXiv Version Number: 1.

\bibitem[Makropoulos et~al., 2018]{makropoulos_review_2018}
Makropoulos, A., Counsell, S.~J., and Rueckert, D. (2018).
\newblock A review on automatic fetal and neonatal brain {MRI} segmentation.
\newblock {\em NeuroImage}, 170:231--248.

\bibitem[{Martin Abadi} et~al., 2015]{martin_abadi_tensorflow_2015}
{Martin Abadi}, {Ashish Agarwal}, {Paul Barham}, {Eugene Brevdo}, {Zhifeng
  Chen}, {Craig Citro}, {Greg S. Corrado}, {Andy Davis}, {Jeffrey Dean},
  {Matthieu Devin}, {Sanjay Ghemawat}, {Ian Goodfellow}, {Andrew Harp},
  {Geoffrey Irving}, {Michael Isard}, {Yangqing Jia}, {Rafal Jozefowicz},
  {Lukasz Kaiser}, {Manjunath Kudlur}, {Josh Levenberg}, {Dandelion Mané},
  {Rajat Monga}, {Sherry Moore}, {Derek Murray}, {Chris Olah}, {Mike Schuster},
  {Jonathon Shlens}, {Benoit Steiner}, {Ilya Sutskever}, {Kunal Talwar}, {Paul
  Tucker}, {Vincent Vanhoucke}, {Vijay Vasudevan}, {Fernanda Viégas}, {Oriol
  Vinyals}, {Pete Warden}, {Martin Wattenberg}, {Martin Wicke}, {Yuan Yu}, and
  {Xiaoqiang Zheng} (2015).
\newblock {TensorFlow}: {Large}-{Scale} {Machine} {Learning} on {Heterogeneous}
  {Systems}.

\bibitem[Payette et~al., 2021]{payette_automatic_2021}
Payette, K., de~Dumast, P., Kebiri, H., Ezhov, I., Paetzold, J.~C., Shit, S.,
  Iqbal, A., Khan, R., Kottke, R., Grehten, P., Ji, H., Lanczi, L., Nagy, M.,
  Beresova, M., Nguyen, T.~D., Natalucci, G., Karayannis, T., Menze, B.,
  Bach~Cuadra, M., and Jakab, A. (2021).
\newblock An automatic multi-tissue human fetal brain segmentation benchmark
  using the {Fetal} {Tissue} {Annotation} {Dataset}.
\newblock {\em Scientific Data}, 8(1):167.

\bibitem[Payette et~al., 2022]{payette_fetal_2022}
Payette, K., Li, H., de~Dumast, P., Licandro, R., Ji, H., Siddiquee, M. M.~R.,
  Xu, D., Myronenko, A., Liu, H., Pei, Y., Wang, L., Peng, Y., Xie, J., Zhang,
  H., Dong, G., Fu, H., Wang, G., Rieu, Z., Kim, D., Kim, H.~G., Karimi, D.,
  Gholipour, A., Torres, H.~R., Oliveira, B., Vilaça, J.~L., Lin, Y.,
  Avisdris, N., Ben-Zvi, O., Bashat, D.~B., Fidon, L., Aertsen, M.,
  Vercauteren, T., Sobotka, D., Langs, G., Alenyà, M., Villanueva, M.~I.,
  Camara, O., Fadida, B.~S., Joskowicz, L., Weibin, L., Yi, L., Xuesong, L.,
  Mazher, M., Qayyum, A., Puig, D., Kebiri, H., Zhang, Z., Xu, X., Wu, D.,
  Liao, K., Wu, Y., Chen, J., Xu, Y., Zhao, L., Vasung, L., Menze, B., Cuadra,
  M.~B., and Jakab, A. (2022).
\newblock Fetal {Brain} {Tissue} {Annotation} and {Segmentation} {Challenge}
  {Results}.
\newblock Publisher: arXiv Version Number: 1.

\bibitem[Pier et~al., 2016]{pier_3d_2016}
Pier, D.~B., Gholipour, A., Afacan, O., Velasco-Annis, C., Clancy, S., Kapur,
  K., Estroff, J.~A., and Warfield, S.~K. (2016).
\newblock {3D} {Super}-{Resolution} {Motion}-{Corrected} {MRI}: {Validation} of
  {Fetal} {Posterior} {Fossa} {Measurements}.
\newblock {\em Journal of Neuroimaging: Official Journal of the American
  Society of Neuroimaging}, 26(5):539--544.

\bibitem[Prayer et~al., 2017]{prayer_isuog_2017}
Prayer, D., Malinger, G., Brugger, P.~C., Cassady, C., De~Catte, L.,
  De~Keersmaecker, B., Fernandes, G.~L., Glanc, P., Gonçalves, L.~F., Gruber,
  G.~M., Laifer-Narin, S., Lee, W., Millischer, A.-E., Molho, M., Neelavalli,
  J., Platt, L., Pugash, D., Ramaekers, P., Salomon, L.~J., Sanz, M.,
  Timor-Tritsch, I.~E., Tutschek, B., Twickler, D., Weber, M., Ximenes, R., and
  Raine-Fenning, N. (2017).
\newblock {ISUOG} {Practice} {Guidelines}: performance of fetal magnetic
  resonance imaging.
\newblock {\em Ultrasound in Obstetrics \& Gynecology}, 49(5):671--680.

\bibitem[Pérez-García et~al., 2021]{perez-garcia_torchio_2021}
Pérez-García, F., Sparks, R., and Ourselin, S. (2021).
\newblock {TorchIO}: {A} {Python} library for efficient loading, preprocessing,
  augmentation and patch-based sampling of medical images in deep learning.
\newblock {\em Computer Methods and Programs in Biomedicine}, 208:106236.

\bibitem[Rajagopalan et~al., 2011]{rajagopalan_local_2011}
Rajagopalan, V., Scott, J., Habas, P.~A., Kim, K., Corbett-Detig, J., Rousseau,
  F., Barkovich, A.~J., Glenn, O.~A., and Studholme, C. (2011).
\newblock Local tissue growth patterns underlying normal fetal human brain
  gyrification quantified in utero.
\newblock {\em The Journal of Neuroscience: The Official Journal of the Society
  for Neuroscience}, 31(8):2878--2887.

\bibitem[Ronneberger et~al., 2015]{navab_u-net_2015}
Ronneberger, O., Fischer, P., and Brox, T. (2015).
\newblock U-{Net}: {Convolutional} {Networks} for {Biomedical} {Image}
  {Segmentation}.
\newblock In Navab, N., Hornegger, J., Wells, W.~M., and Frangi, A.~F.,
  editors, {\em Medical {Image} {Computing} and {Computer}-{Assisted}
  {Intervention} – {MICCAI} 2015}, volume 9351, pages 234--241. Springer
  International Publishing, Cham.
\newblock Series Title: Lecture Notes in Computer Science.

\bibitem[Rote and Vegter, 2006]{boissonnat_computational_2006}
Rote, G. and Vegter, G. (2006).
\newblock Computational {Topology}: {An} {Introduction}.
\newblock In Boissonnat, J.-D. and Teillaud, M., editors, {\em Effective
  {Computational} {Geometry} for {Curves} and {Surfaces}}, pages 277--312.
  Springer Berlin Heidelberg.

\bibitem[Rousseau et~al., 2006]{rousseau_registration-based_2006}
Rousseau, F., Glenn, O.~A., Iordanova, B., Rodriguez-Carranza, C., Vigneron,
  D.~B., Barkovich, J.~A., and Studholme, C. (2006).
\newblock Registration-{Based} {Approach} for {Reconstruction} of
  {High}-{Resolution} {In} {Utero} {Fetal} {MR} {Brain} {Images}.
\newblock {\em Academic Radiology}, 13(9):1072--1081.

\bibitem[Salomon et~al., 2006]{salomon_third-trimester_2006}
Salomon, L., Ouahba, J., Delezoide, A.-L., Vuillard, E., Oury, J.-F., Sebag,
  G., and Garel, C. (2006).
\newblock Third-trimester fetal {MRI} in isolated 10- to 12-mm
  ventriculomegaly: is it worth it?
\newblock {\em BJOG: An International Journal of Obstetrics and Gynaecology},
  113(8):942--947.

\bibitem[Severino et~al., 2020]{severino_definitions_2020}
Severino, M., Geraldo, A.~F., Utz, N., Tortora, D., Pogledic, I., Klonowski,
  W., Triulzi, F., Arrigoni, F., Mankad, K., Leventer, R.~J., Mancini, G.
  M.~S., Barkovich, J.~A., Lequin, M.~H., and Rossi, A. (2020).
\newblock Definitions and classification of malformations of cortical
  development: practical guidelines.
\newblock {\em Brain}, 143(10):2874--2894.

\bibitem[Tarui et~al., 2018]{tarui_disorganized_2018}
Tarui, T., Madan, N., Farhat, N., Kitano, R., Ceren~Tanritanir, A., Graham, G.,
  Gagoski, B., Craig, A., Rollins, C.~K., Ortinau, C., Iyer, V., Pienaar, R.,
  Bianchi, D.~W., Grant, P.~E., and Im, K. (2018).
\newblock Disorganized {Patterns} of {Sulcal} {Position} in {Fetal} {Brains}
  with {Agenesis} of {Corpus} {Callosum}.
\newblock {\em Cerebral Cortex (New York, N.Y.: 1991)}, 28(9):3192--3203.

\bibitem[Tierney and Nelson, 2009]{tierney_brain_2009}
Tierney, A.~L. and Nelson, C.~A. (2009).
\newblock Brain {Development} and the {Role} of {Experience} in the {Early}
  {Years}.
\newblock {\em Zero to Three}, 30(2):9--13.

\bibitem[Tourbier et~al.,
  2019]{tourbier_sebastien_sebastientourbiermialsuperresolutiontoolkit_2019}
Tourbier, S., Bresson, X., Hagmann, P., Meuli, R., and Bach~Cuadra, M. (2019).
\newblock sebastientourbier/mialsuperresolutiontoolkit: {MIAL}
  {Super}-{Resolution} {Toolkit} v1.0.

\bibitem[Tourbier et~al., 2015]{tourbier_efficient_2015}
Tourbier, S., Bresson, X., Hagmann, P., Thiran, J.-P., Meuli, R., and Cuadra,
  M.~B. (2015).
\newblock An efficient total variation algorithm for super-resolution in fetal
  brain {MRI} with adaptive regularization.
\newblock {\em NeuroImage}, 118:584--597.

\bibitem[Uus et~al., 2022]{uus_retrospective_2022}
Uus, A.~U., Egloff~Collado, A., Roberts, T.~A., Hajnal, J.~V., Rutherford,
  M.~A., and Deprez, M. (2022).
\newblock Retrospective motion correction in foetal {MRI} for clinical
  applications: existing methods, applications and integration into clinical
  practice.
\newblock {\em The British Journal of Radiology}, page 20220071.

\bibitem[Vasung et~al., 2016]{vasung_quantitative_2016}
Vasung, L., Lepage, C., Radoš, M., Pletikos, M., Goldman, J.~S., Richiardi,
  J., Raguž, M., Fischi-Gómez, E., Karama, S., Huppi, P.~S., Evans, A.~C.,
  and Kostovic, I. (2016).
\newblock Quantitative and {Qualitative} {Analysis} of {Transient} {Fetal}
  {Compartments} during {Prenatal} {Human} {Brain} {Development}.
\newblock {\em Frontiers in Neuroanatomy}, 10.

\bibitem[Vasung et~al., 2020]{vasung_spatiotemporal_2020}
Vasung, L., Rollins, C.~K., Velasco-Annis, C., Yun, H.~J., Zhang, J., Warfield,
  S.~K., Feldman, H.~A., Gholipour, A., and Grant, P.~E. (2020).
\newblock Spatiotemporal {Differences} in the {Regional} {Cortical} {Plate} and
  {Subplate} {Volume} {Growth} during {Fetal} {Development}.
\newblock {\em Cerebral Cortex}, 30(8):4438--4453.

\bibitem[Wright et~al., 2014]{wright_automatic_2014}
Wright, R., Kyriakopoulou, V., Ledig, C., Rutherford, M., Hajnal, J., Rueckert,
  D., and Aljabar, P. (2014).
\newblock Automatic quantification of normal cortical folding patterns from
  fetal brain {MRI}.
\newblock {\em NeuroImage}, 91:21--32.

\bibitem[Xia et~al., 2019]{xia_fetal_2019}
Xia, J., Wang, F., Benkarim, O.~M., Sanroma, G., Piella, G.,
  González~Ballester, M.~A., Hahner, N., Eixarch, E., Zhang, C., Shen, D., and
  Li, G. (2019).
\newblock Fetal cortical surface atlas parcellation based on growth patterns.
\newblock {\em Human Brain Mapping}, 40(13):3881--3899.

\bibitem[Yeghiazaryan and Voiculescu, 2018]{yeghiazaryan_family_2018}
Yeghiazaryan, V. and Voiculescu, I. (2018).
\newblock Family of boundary overlap metrics for the evaluation of medical
  image segmentation.
\newblock {\em Journal of Medical Imaging}, 5(01):1.

\bibitem[Yushkevich et~al., 2006]{yushkevich_user-guided_2006}
Yushkevich, P.~A., Piven, J., Hazlett, H.~C., Smith, R.~G., Ho, S., Gee, J.~C.,
  and Gerig, G. (2006).
\newblock User-guided {3D} active contour segmentation of anatomical
  structures: {Significantly} improved efficiency and reliability.
\newblock {\em NeuroImage}, 31(3):1116--1128.

\end{thebibliography}


\end{document}